\documentclass[twocolumn,superscriptaddress,aps,prl]{revtex4-1}
\usepackage{amsmath,amssymb,color,graphicx,dcolumn,bm}

\usepackage{lineno}
\usepackage{dsfont}
\usepackage{comment}
\usepackage{color,soul,float,physics}

\begin{document}

\title{Quantum Metrology with Strongly Interacting Spin Systems}

\affiliation{Department of Physics, Harvard University, Cambridge, Massachusetts 02138, USA}
\affiliation{School of Engineering and Applied Sciences, Harvard University, Cambridge, Massachusetts 02138, USA}
\affiliation{Department of Physics, University of California Berkeley, Berkeley, California 94720, USA}
\affiliation{Tsukuba Research Center for Energy Materials Science, Faculty of Pure and Applied Sciences, University of Tsukuba, Tsukuba, Ibaraki 305-8573 Japan}
\affiliation{Institut f{\"u}r Quantenoptik, Universit{\"a}t Ulm, 89081 Ulm, Germany}
\affiliation{Takasaki Advanced Radiation Research Institute, 1233 Watanuki, Takasaki, Gunma 370-1292, Japan}
\affiliation{Sumitomo Electric Industries Ltd., Itami, Hyougo, 664-0016, Japan}
\affiliation{Research Laboratory of Electronics and Department of Nuclear Science and Engineering, Massachusetts Institute of Technology, Cambridge, Massachusetts 02139, USA}
\affiliation{Department of Chemistry and Chemical Biology, Harvard University, Cambridge, Massachusetts 02138, USA}

\author{
Hengyun Zhou$^{1,*}$, Joonhee Choi$^{1,2,*}$, Soonwon Choi$^3$, Renate Landig$^{1}$, Alexander M. Douglas$^1$, Junichi Isoya$^{4}$, Fedor Jelezko$^{5}$, Shinobu Onoda$^{6}$, Hitoshi Sumiya$^{7}$, Paola Cappellaro$^8$, Helena S. Knowles$^1$, Hongkun Park$^{1,9}$ and Mikhail D. Lukin$^{1,\dagger}$
}

\begin{abstract}
Quantum metrology is a powerful tool for explorations of fundamental physical phenomena~\cite{demille2017probing} and applications in material science~\cite{casola2018probing} and biochemical analysis~\cite{aslam2017nanoscale,glenn2018high}. While in principle the sensitivity can be improved by increasing the density of sensing particles, in practice this improvement is severely hindered by interactions between them~\cite{mitchell2019sensor}. Here, using a dense ensemble of interacting electronic spins in diamond, we demonstrate a novel approach to quantum metrology to surpass such limitations. It is based on a new method of robust quantum control, which allows us to simultaneously suppress the undesired effects associated with spin-spin interactions, disorder and control imperfections, enabling a five-fold enhancement in coherence time compared to state-of-the-art control sequences~\cite{degen2017quantum}. Combined with optimal initialization and readout protocols, this allows us to achieve an AC magnetic field sensitivity well beyond the previous limit imposed by interactions, opening a new regime of high-sensitivity solid-state ensemble magnetometers.
\end{abstract}

\maketitle 

\begin{figure}[h]
\includegraphics[width=.48\textwidth]{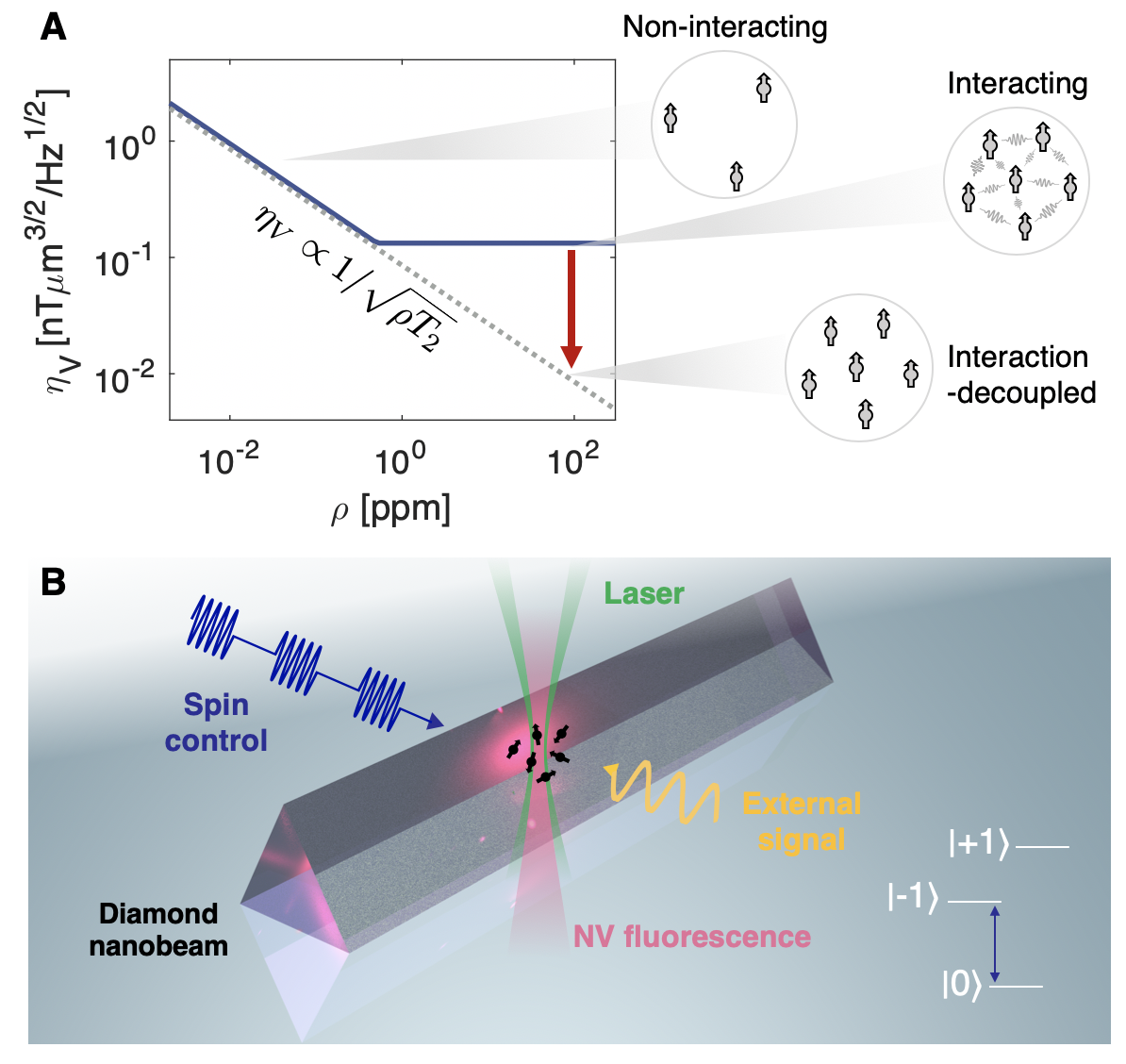}
\caption{{\bf Interaction limit to spin ensemble quantum sensing.} 
{\bf (A)} Volume-normalized magnetic field sensitivity as a function of total spin density. The dashed line denotes the standard quantum limit scaling and the solid curve shows the behaviour when interactions between spins are taken into account for the typical readout efficiency factor $C = 0.01$~\cite{taylor2008high}. The sensitivity plateaus beyond a critical density due to a coherence time reduction. Robust interaction decoupling (red arrow) allows us to break the interaction limit. {\bf (B)} Illustration of the black diamond nanobeam used as a spin ensemble quantum sensor. Microwave and optical excitation are delivered to the NV spins to control and read out their spin states, and an AC magnetic field is used as a target sensing signal. The inset shows the three magnetic sublevels, $\ket{0}$ and $\ket{\pm1}$, in the ground state of NV centers, where two levels, $\ket{0,-1}$ are addressed using resonant microwave driving. All measurements are performed at room temperature under ambient conditions.}
\label{fig:fig1}
\end{figure}
Electronic spins associated with color centers in diamond have recently emerged as a promising platform for nanoscale precision sensing and imaging, with superior sensitivity and spatial resolution~\cite{schirhagl2014nitrogen,rondin2014magnetometry}. A common approach to improving sensitivity of such quantum systems is to utilize a dense ensemble of individual sensors and take advantage of parallel averaging.
However, beyond a certain density, undesired interactions between sensors result in a rapid decay of the ensemble coherence~\cite{mitchell2019sensor}, limiting the overall sensitivity of the quantum system (Fig.~1A). 
Moreover, in practice, disorder and control errors further deteriorate coherence and metrological sensitivity of such interacting many-body quantum systems. 
Over the past few decades, pulsed control techniques realizing dynamical decoupling and motional averaging have been developed and deployed to manipulate interacting ensembles of quantum systems. While extremely successful in the context of nuclear magnetic resonance (NMR), magnetic resonance imaging (MRI)~\cite{haeberlen1968coherent,waugh1968approach,mansfield1971symmetrized,burum1979analysis,chmelka1989some,cory1990time} and atomic gas magnetometers~\cite{allred2002high}, the efficacy of these techniques is severely limited in the presence of strong disorder and other imperfections. Therefore, these techniques are not directly applicable to quantum sensors based on electronic spin ensembles~\cite{SM}, where such effects are prominent.

In this Letter, we demonstrate a new approach for quantum sensing with disordered interacting spin ensembles.
Our method uses periodic pulsed manipulation (Floquet engineering) of a spin ensemble~\cite{haeberlen1968coherent} to detect an external signal of interest with high sensitivity, while simultaneously decoupling the effects of interactions and disorder, and being fault-tolerant against the leading-order imperfections arising from finite pulse durations and experimental control errors. Specifically, we introduce a set of simple rules imposed on the pulse sequence for \textit{disordered, interacting} systems and a new, generalized picture of AC field sensing that is essential in the interacting regime. These go beyond existing dynamical decoupling techniques for \textit{non-interacting} spin systems~\cite{suter2016,degen2017quantum,lang2017enhanced,genov2017arbitrarily} and low-disorder NMR systems~\cite{haeberlen1968coherent,waugh1968approach,mansfield1971symmetrized,burum1979analysis,chmelka1989some,cory1990time}, allowing us to design and implement a sequence that breaks the sensitivity limit on AC sensing imposed by spin-spin interactions for the first time.

Our experimental system consists of a dense electronic spin ensemble of NV centers in diamond~\cite{kucsko2018critical}, as shown in Fig.~\ref{fig:fig1}B. NV centers exhibit long-lived spin coherence even at room temperature and are excellent sensors of magnetic fields, electric fields, pressure, and temperature~\cite{schirhagl2014nitrogen,rondin2014magnetometry,taylor2008high,degen2008scanning}.
Our sample has a high density of NV centers ($\sim$15~ppm, see \cite{SM}), with long-range magnetic dipolar interactions between the spins as well as strong on-site disorder originating from other paramagnetic impurities, inhomogeneous strain in the diamond lattice, and local electric fields~\cite{kucsko2018critical}.
The bulk diamond is etched into a nanobeam to improve control homogeneity and confine the probing volume to $V = 8.1(9)\times 10^{-3}~\mu$m$^3$~\cite{SM}.
The NV center ground state is an electronic $S=1$ spin, and we apply a static magnetic field to isolate an ensemble of effective two-level systems formed of NVs with the same crystallographic orientation.
We initialize and detect the spin states optically and use resonant microwave excitation to drive coherent spin dynamics. 
See~\cite{SM} for further details of the measurement sequences.

\begin{figure*}
  \begin{minipage}[c]{0.48\textwidth}
    \includegraphics[width=\textwidth]{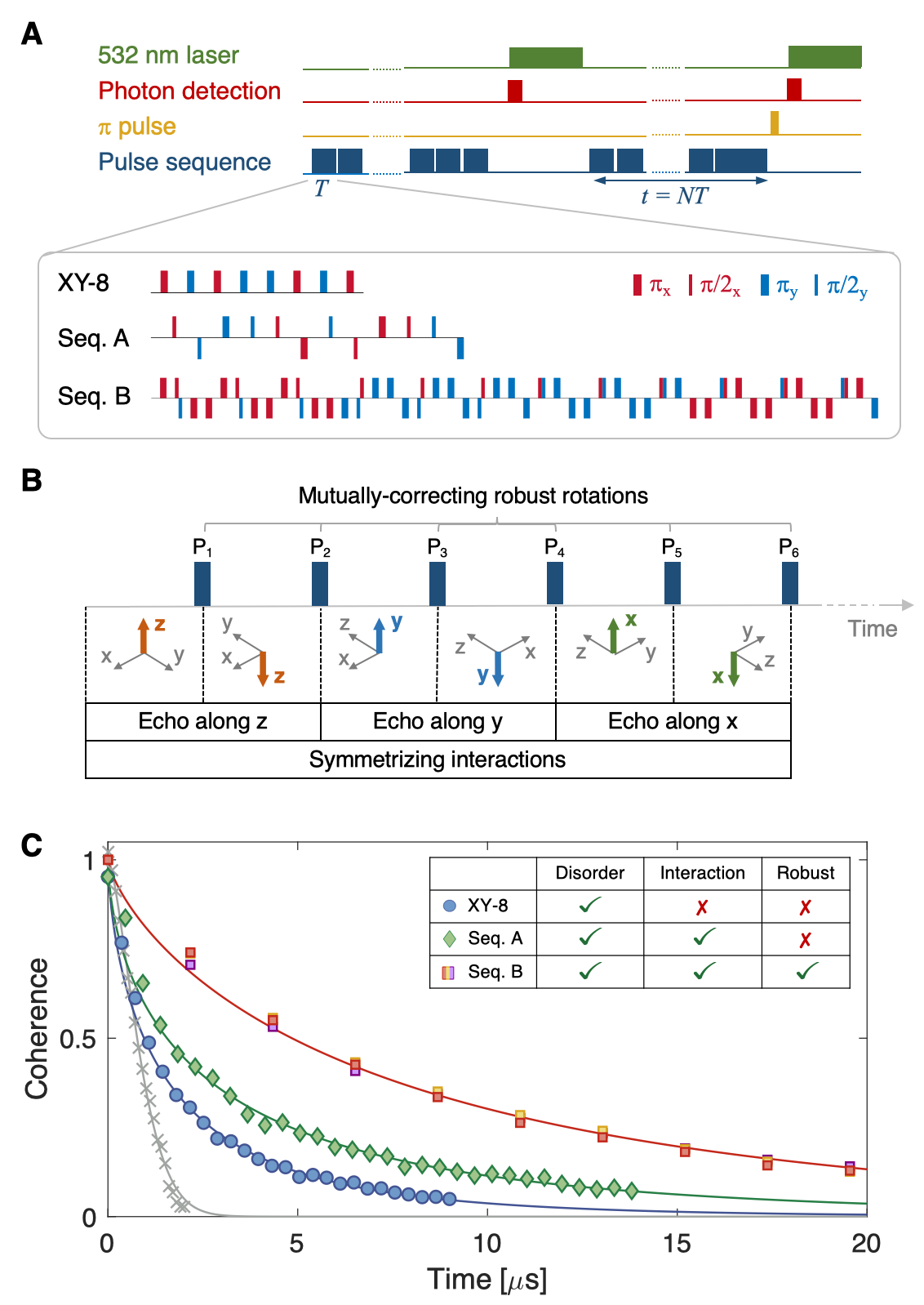}
  \end{minipage}\hfill
  \begin{minipage}[c]{0.48\textwidth}
    \caption{{\bf Robust dynamical decoupling.} {\bf (A)} Measurement protocol. NVs are initialized using pulsed laser excitation at 532 nm (green trace) and read out through emitted photons detected by a single photon counting module (red trace). We perform $N$ repetitions of a sensing sequence unit of length $T$ (blue trace) and repeat the same measurement with an additional $\pi$ pulse (yellow trace) acting on the NV centers for differential readout of the spin state~\cite{SM}. The box illustrates the details of different pulse sequences, XY-8, Seq.~A and Seq.~B, composed of $\pi/2$ and $\pi$ rotations along $\hat{x}$ and $\hat{y}$ axes. Bars above (below) the line indicate driving along positive (negative) axis directions.
{\bf (B)} Key concepts for sequence design. The sequence is described by pulses $P_i$ and the time-dependent frame transformation of the system between the pulses. We highlight the orientation of each rotated frame by the axis that points along the $\hat{z}$ axis of the fixed external reference frame.
Decoupling sequences are designed by imposing average Hamiltonian conditions on the evolution of the highlighted axis. For example, the effects of disorder can be cancelled by implementing an echo-like evolution, $+\hat{\mu} \rightarrow -\hat{\mu}$ where $\mu = x,y,z$ (top row), and interactions are symmetrized by equal evolution in each of the $\hat{x},\hat{y}$ and $\hat{z}$ axes in the transformed frames (bottom row). Additionally, the pulse sequence is designed to mutually correct rotation angle errors and finite pulse duration effects~\cite{SM}.
{\bf (C)} Experimental performance of different sequences with their respective decoupling features (inset). We fit the decoherence profile with a stretched exponential $e^{-(t/T_2)^\alpha}$ (solid curves) to extract the coherence time $T_2$ for each sequence~\cite{SM}. A simple spin-echo (gray crosses), XY-8 (blue circles) and Seq.~A (green diamonds) show $T_2 = 0.98(2)~\mu$s, $1.6(1)~\mu$s, and $2.8(1)~\mu$s with $\alpha = 1.5(1), 0.66(2),$ and $0.61(3)$, respectively. Seq.~B (squares), designed to correct for all leading-order effects of interactions, disorder and control imperfections, gives $T_2 = 7.9(2)~\mu$s with $\alpha = 0.75(2)$. We confirm that its coherence time is independent of the initial state prepared along $\hat{x}, \hat{y}$ and $\hat{z}$ axes, as shown in red, yellow and purple, respectively. All sequences have pulse spacing $\tau=25~$ns and $\pi$-pulse width $\tau_{\pi}=20~$ns.
} 
  \end{minipage}
  \label{fig:fig2}
\end{figure*}

Spin echo measurements~\cite{slichter2013principles} reveal that the coherence decay time $T_2$ in our dense NV ensemble is limited to only $1.0~\mu s$ (Fig.~\ref{fig:fig2}C, gray crosses).
The conventional method to extend $T_2$ beyond the spin echo is the XY-8 dynamical decoupling sequence~\cite{suter2016}, consisting of equally-spaced $\pi$ pulses along the $\hat{x}$ and $\hat{y}$ axes (Fig.~\ref{fig:fig2}A, top row). In our system, however, the XY-8 sequence only provides a small improvement ($T_2 = 1.6~\mu$s, Fig.~\ref{fig:fig2}C, blue circles), since the XY-8 $T_2$ is limited by strong spin-spin interactions, which are not affected by $\pi$ rotations (see~\cite{SM} for density scaling measurements which further confirm that the XY-8 $T_2$ is interaction-limited).

In order to significantly extend $T_2$ in the presence of interactions and control imperfections, we use a novel approach to design pulse sequences~\cite{choi2019robust}. We model our spin system, including the control fields and the external AC magnetic field target signal, by the Hamiltonian~\cite{SM,kucsko2018critical}
\begin{align}
	H = H_s + H_\Omega(t) + H_{\text{AC}}(t),
\end{align}
where the internal system Hamiltonian is $H_s = \sum_i h_i S_i^z + \sum_{ij}\frac{J_{ij}}{r_{ij}^3}(S_i^x S _j^x+S_i^y S_j^y-S_i^zS_j^z)$, global spin-control pulses are given by $H_\Omega (t) = \sum_i (\Omega_i^x(t)S_i^x+\Omega_i^y(t)S_i^y) $ and the external target signal is $H_{\text{AC}}(t) = \gamma_\text{NV} B_{\text{AC}} \cos(2\pi f_\text{AC} t - \phi) \sum_i S_i^z$. Here, $S_i^\mu$ ($\mu=x,y,z$) are spin-1/2 operators, $h_i$ is a random on-site disorder potential with zero mean that follows a Gaussian distribution with standard deviation $W=(2\pi)~4.0$~MHz, $J_{ij}/r_{ij}^3$ is the anisotropic dipolar interaction strength between two spins of the same crystallographic orientation at a distance $r_{ij}$, with strength $J =(2\pi)~35$~kHz at a typical separation of 11 nm, $\Omega_i^{x,y}(t)$ are the global control amplitudes exhibiting weak position dependence due to spatial field inhomogeneities, $\gamma_\text{NV}$ is the gyromagnetic ratio of the NV center, and $B_{\text{AC}}, f_\text{AC}$ and $\phi$ are the amplitude, frequency and phase of the target AC signal, respectively.

Our approach employs average Hamiltonian theory to engineer the system evolution through pulsed periodic manipulation of the spins~\cite{haeberlen1968coherent,slichter2013principles}.
A sequence composed of $n$ equidistant control pulses \{$P_k; k = 1,2,.., n$\} with spacing $\tau$ defines a unitary time-evolution operator $\mathcal{U}(T) = P_k e^{-iH_s \tau} \cdots P_1 e^{-iH_s \tau}$ over the Floquet period $T$.
If the pulse spacing $\tau$ is much shorter than the timescales of the system Hamiltonian ($\tau \ll \frac{1}{W}, \frac{1}{J}$), the unitary operator $\mathcal{U}(T)$ can effectively be approximated by a time-independent average Hamiltonian as $\mathcal{U}(T) \approx  e^{-iH_{\text{avg}} T}$, with $H_{\text{avg}} = \frac{1}{T}\sum_{k=1}^n \tilde{H}_k$, and $\tilde{H}_k = (P_{k-1} \cdots P_1)^\dagger H_s (P_{k-1} \cdots P_1)$, $\tilde{H}_1=H_s$.
Motivated by this picture, we develop a pulse sequence which generates a desirable form of $H_{\text{avg}}$ from the $H_s$ intrinsic to the system.

Hamiltonian engineering can be understood as the result of a sequence of frame transformations (also known as toggling frame transformations) which rotate the spin operators in the interaction picture (Fig.~\ref{fig:fig2}B): for example, a $\pi$-pulse flips $S_i^{z} \rightarrow -S_i^{z}$, while a $\pi/2$-pulse rotates $S_i^z \rightarrow \pm S_i^{x,y}$ depending on the rotation axis. 
Importantly, the average Hamiltonian is uniquely specified by such toggling frame transformations of the $S^z$ operator~\cite{SM}, resulting in simple decoupling conditions that facilitate the procedure to find desired pulse sequences.
\twocolumngrid
For example, any pulse sequence in which the transformed $S^z$ operator spends equal time along the positive and negative direction for each axis---effectively producing a spin echo along all three axes---suppresses the on-site disorder Hamiltonian.
Similarly, in order to symmetrize the dipolar interaction into a Heisenberg interaction Hamiltonian (where polarized states are eigenstates and coherence is preserved~\cite{choi2017dynamical}), we require that the transformed $S^z$ operator spend an equal amount of time in each direction $\hat{x}$, $\hat{y}$, $\hat{z}$~\cite{SM,waugh1968approach}.
Furthermore, we can prioritize one condition over another to find a pulse sequence that better suits a given system; since disorder is dominant in our spin ensemble ($W\gg J$), we perform the echo operation more frequently than interaction symmetrization.

In realistic situations, the above strategy will be affected by various imperfections, such as disorder and interactions acting during the finite pulse durations and errors of each control pulse, resulting in imperfections $\delta H_\text{avg}$ to the target effective Hamiltonian, $H_\text{eff} = H_\text{avg} + \delta H_\text{avg}$. Indeed, if we simply design a pulse sequence that decouples disorder and interactions only in the {\it ideal} pulse limit (Seq.~A in Fig.~2A), we see only a marginal increase in coherence time compared to XY-8, yielding $T_2 = 2.8~\mu$s (Fig.~2C, green diamonds). A careful examination of Seq.~A reveals that pulse-related imperfections play a dominant role in the dynamics, illustrating the importance of robust sequence design ~\cite{SM}.

To address this key challenge, a number of strategies have been proposed that aim to cancel or control the undesired Hamiltonian terms acting during the finite pulse duration~\cite{mansfield1971symmetrized,burum1979analysis,chmelka1989some,cory1990time,lang2017enhanced,genov2017arbitrarily}; however, a simple and systematic approach to treating the imperfections in a general setting, particularly for \textit{interacting} systems, is still lacking. Remarkably, we find that the transformations of the $S^z$ operator during the \textit{free evolution intervals} are in fact sufficient to predict and suppress the errors during pulse rotations. This is a consequence of the fact that the unwanted residual Hamiltonian acting within the finite pulse duration can be uniquely determined by the two toggling-frame Hamiltonians on either side of the control pulse~\cite{SM}. This insight motivates us to describe the $S^z$ operator transformations using the matrix, ${\bf F} = [F_{\mu,k}] = 2\textrm{Tr}[S^\mu \tilde{S}^z_k]$, $\mu = \hat{x},\hat{y},\hat{z}$, where $\tilde{S}^z_k$ is the transformed spin operator within the $k$-th free evolution period. Crucially, this allows us to construct a simple set of algebraic conditions imposed on the matrix ${\bf F}$ to formalize the above decoupling rules, enabling not only the suppression of disorder and interaction effects during free evolution periods, but also the cancellation of dominant imperfections arising from finite pulses as well as rotation angle errors (see Eqs.~(S8-S11) in \cite{SM} for the exact expressions of decoupling conditions). This also enables system-customized design for optimal AC-field sensing, taking into account the energetic hierarchy between on-site disorder, spin-spin interactions and control imperfections in a given system, significantly extending beyond existing techniques.

By satisfying these rules, we can thus systematically generate robust pulse sequences that to first order yield a pure Heisenberg Hamiltonian with $\delta H_\text{avg}=0$. Seq.~B, as shown in Fig.~\ref{fig:fig2}A, is an example of a robust pulse sequence designed with our formalism~\cite{SM}. Due to its robustness against all leading-order effects, it shows a significant extension of coherence time compared to the sequences described above, reaching $T_2 =7.9$ $\mu$s (Fig.~\ref{fig:fig2}C, squares). Moreover, the coherence time is independent of the initial state.

We now apply this method to quantum sensing, where our goal is to robustly engineer the dynamics of the spin ensemble to be sensitive to the target sensing signal.
AC magnetic field sensing typically uses periodic inversions of the spin operator between $S^z$ and $-S^z$ in the interaction picture, driven by a train of equidistant $\pi$ pulses at a separation of $\frac{1}{2f_{\mathrm{AC}}}$. 
This modulation causes cumulative precession of the sensor spin when the AC field sign change coincides with the frame inversion, resulting in high sensitivity to a signal field at $f_{\mathrm{AC}}$.
Our interaction-decoupling sequences explore all three frame directions $S^x$, $S^y$, $S^z$ and AC selectivity requires synchronized periodic frame inversions in each of the three axes, while preserving the desired $H_{\text{avg}}$ and suppressing $\delta H_\text{avg}$ to maintain long coherence times.

\begin{figure}[H]
\begin{center}
\includegraphics[width=.48\textwidth]{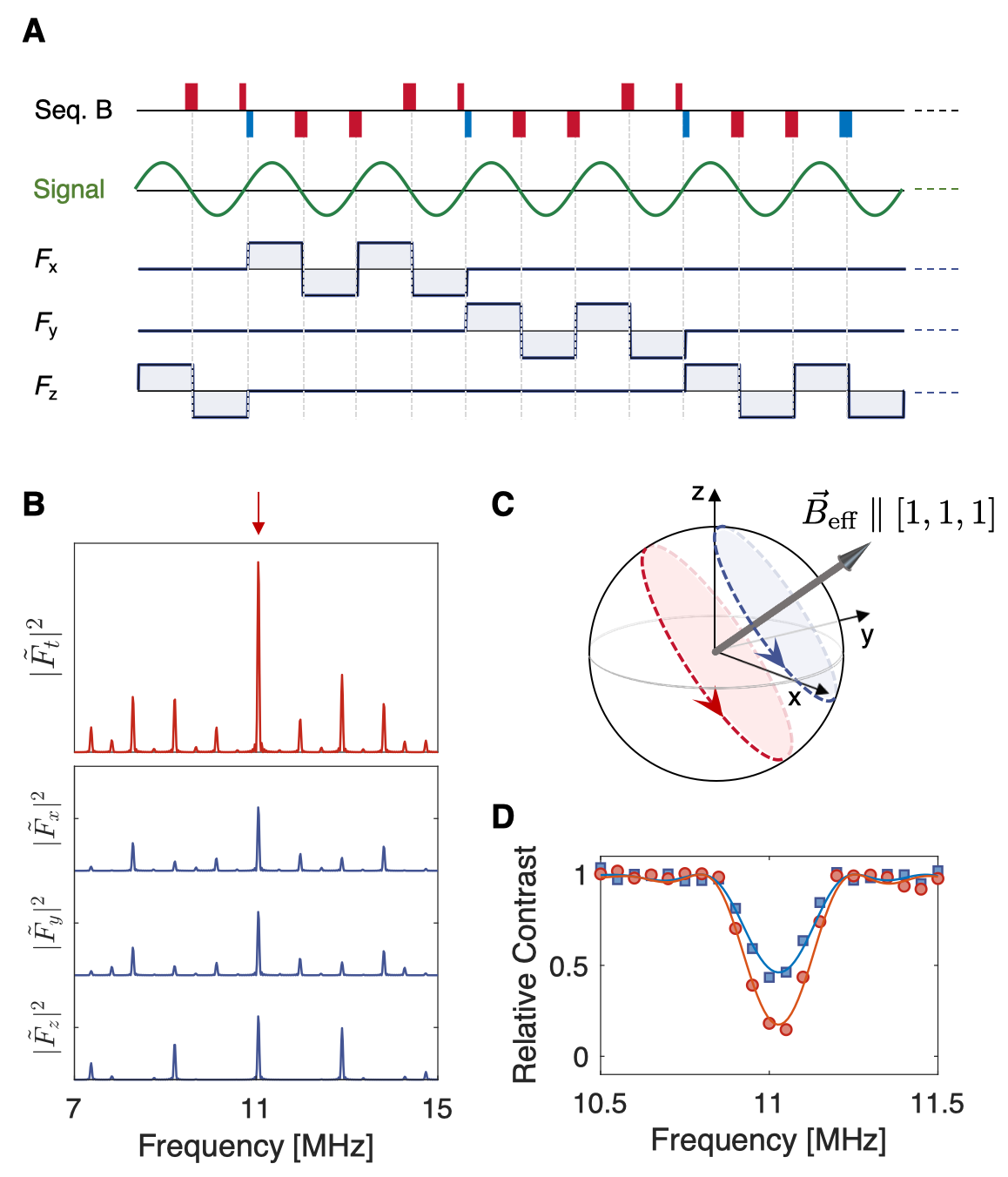}
\end{center}
\caption{{\bf Optimal sensing with unconventional spin state preparation. } 
{\bf (A)}  Pulse sequence (top row) and three-axis time-domain modulation functions (blue solid curves) for the first 14 free evolution times of Seq.~B. Red/blue bars in Seq.~B indicate rotation pulses as defined in Fig.~2A. The modulation period along each axis is synchronized to an AC sensing signal (green curve).
{\bf (B)} Frequency-domain modulation function $|\tilde{F}_{x,y,z}(f)|^2$ for Seq.~B, with pulse spacing $\tau=25$ ns and $\pi$-pulse width $\tau_{\pi}=20$ ns. The total strength $|\tilde{F}_t(f)|^2$ is obtained from individual axis amplitudes $\tilde{F}_{x,y,z}(f)$ considering their relative phases in the frequency domain~\cite{SM}. The principal resonance is highlighted by a red arrow, yielding maximum sensitivity. 
{\bf (C)} Illustration of the effective magnetic field created by Seq.~B at the principal resonance. In the average Hamiltonian picture, the $\hat{z}$-direction sensing field in the external reference frame transforms to the [1,1,1]-direction field $\vec{B}_\text{eff}$ in the effective spin frame with a reduced strength~\cite{SM}. Optimal sensitivity is achieved by initializing the spins into the plane perpendicular to the effective magnetic field direction. This optimal state preparation allows the spins to precess along the trajectory with the largest contrast (red dashed line). For comparison, the precession evolution for initialization to the conventional $\hat{x}$ axis is shown  as a blue dashed line.
{\bf (D)} Sensing resonance spectra near the principal resonance. The optimal initialization (red) shows greater contrast than the $\hat{x}$-axis initialization (blue). Markers indicate experimental data and solid lines denote theoretical predictions calculated from the frequency-domain modulation functions~\cite{SM}.
}
\label{fig:fig3}
\end{figure}

In Fig.~\ref{fig:fig3}A we illustrate how this is achieved in Seq.~B.
The pulses lead to periodic changes in the sign and orientation of the interaction-picture $S^z$ operator, depicted by the time-domain modulation functions for each axis direction, $F_x, F_y$ and $F_z$.
The detailed resonance properties of the pulse sequence can be characterized by the Fourier transforms $\tilde{F}_\mu(f) = |\tilde{F}_\mu(f)|e^{-i\tilde{\phi}_\mu(f)}$ of $F_\mu$ for $\mu = x,y,z$, where $\tilde{\phi}_\mu(f)$ is the spectral phase for a given axis $\mu$. 
Fig.~\ref{fig:fig3}B shows the calculated spectral intensities along different axes, $|\tilde{F}_x(f)|^2$,$|\tilde{F}_y(f)|^2$, and $|\tilde{F}_z(f)|^2$, as well as the total intensity $|\tilde{F}_t(f)|^2$~\cite{SM}.
At the dominant resonance of the total intensity (red arrow in Fig.~\ref{fig:fig3}B), all three axes exhibit a phase-locked periodic sign modulation (Fig.~\ref{fig:fig3}A), leading to constructive phase accumulation and high sensitivity.

In order to intuitively understand our sensing protocol and quantify its sensitivity, we generalize the average Hamiltonian analysis to incorporate AC signal fields, finding~\cite{SM}
\begin{align}
H_\text{avg,AC} &= \gamma_\text{NV} B_\text{AC} \sum_i \text{Re}\left[\sum_{\mu=x,y,z} \tilde{F}_\mu(f_\text{AC})S_i^\mu e^{i\phi} \right]\nonumber\\&=\gamma_\text{NV} \vec{B}_\text{eff} \cdot \sum_i \vec{S}_i,
\end{align}
where $\vec{B}_\text{eff}$ is an effective magnetic field vector in the interaction picture which appears static to the driven spins.
This allows a simple interpretation of our scheme: the spins undergo a precession around $\vec{B}_\text{eff}$, with the field orientation and magnitude determined by the frequency-domain modulation functions $\tilde{F}_\mu$ and $|\tilde{F}_t|$, respectively.
For conventional sequences with $\vec{B}_\text{eff}\parallel \hat{z}$, the requirements for interaction symmetrization dictate that only $1/3$ of the total sensing time can be spent along any given axis, resulting in a significant loss of sensitivity.
In contrast, for our optimal Seq.~B, the signal at the principal resonance $f_\text{AC}$ gives rise to $|\tilde{F}_x| = |\tilde{F}_y| = |\tilde{F}_z|$ with $\tilde{\phi}_x = \tilde{\phi}_y = \tilde{\phi}_z$, leading to $\vec{B}_\text{eff} \propto [\frac{1}{3},\frac{1}{3},\frac{1}{3}]$ with the total strength $|\vec{B}_\text{eff}|$ reduced by a factor of only $1/\sqrt{3}$.
This reduction is fundamental and, in fact, close to optimal~\cite{choi2019robust}, given the requirements to suppress the effects of spin-spin interactions via symmetrization.
However, despite the reduction, the sensitivity is still improved because the coherence time is extended by interaction suppression.
Moreover, the sensitivity to the signal $B_\text{AC}$ is maximized when the spins are initialized perpendicular to the $\vec{B}_\text{eff}$ direction---to allow the largest precession orbit (Fig.~\ref{fig:fig3}C)---and the corresponding optimal readout requires an unconventional rotation axis $[-1,1,0]$ and angle $\arccos (\sqrt{2/3})$ to bring the precession plane parallel to the $\hat{z}$ axis.
Indeed, as shown in Fig.~\ref{fig:fig3}D, we observe a larger contrast when spins are initialized in an optimal direction along $[1,1,-2]$, compared to an initialization in the conventional $\hat{x}$ direction.

We now proceed to characterize the AC magnetic field sensitivity, defined as the minimum detectable signal amplitude per unit time, $\eta = \frac{\sigma_S}{|dS/dB_\text{AC}|}$. Here, $S$ is the spin contrast, $\sigma_S$ is the uncertainty of $S$ for one second of averaging, and $|dS/dB_\text{AC}|$ is the gradient of $S$ with respect to the field amplitude $B_\text{AC}$.

\begin{figure}[h]
\begin{center}
\includegraphics[width=.48\textwidth]{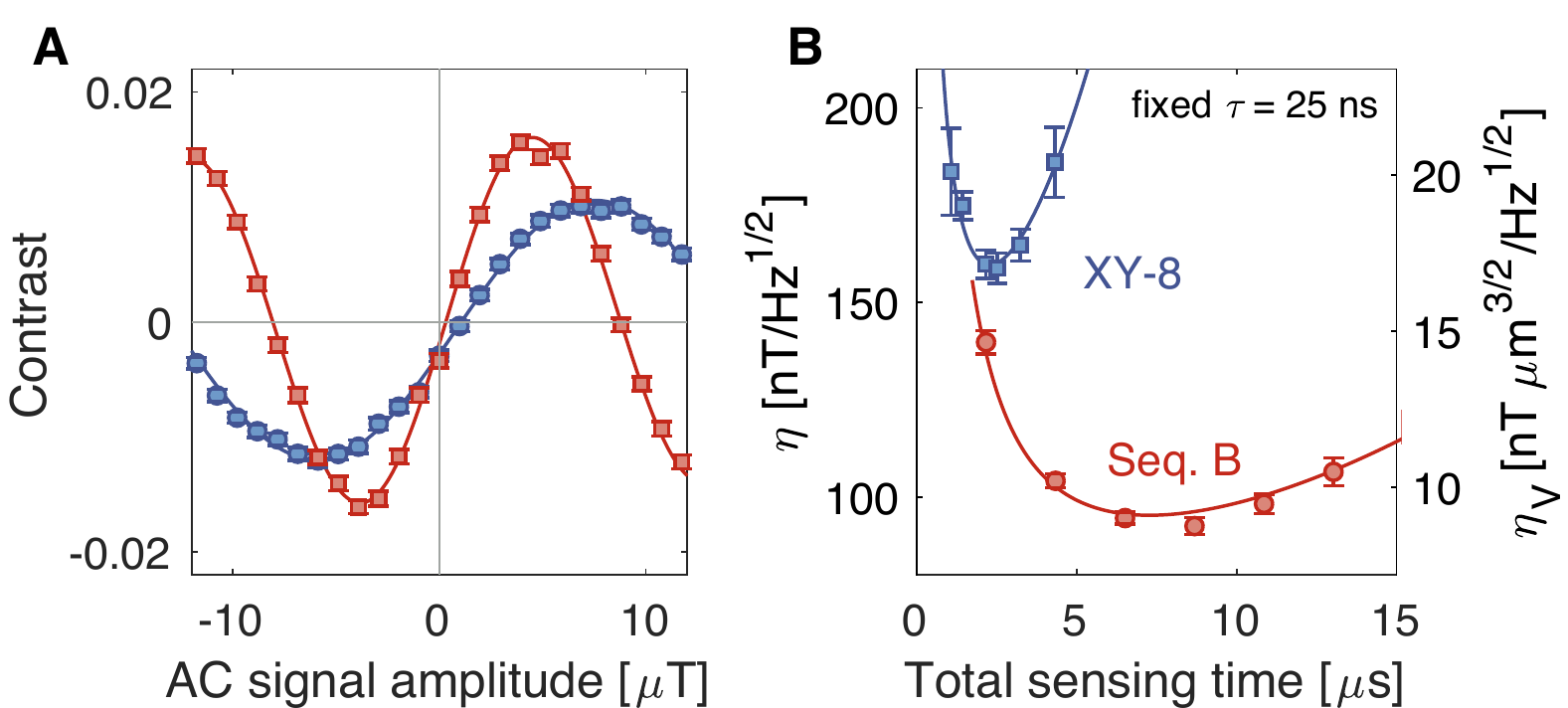}
\end{center}
\caption{{\bf Demonstration of sensitivity enhancement.} 
{\bf (A)} Observed spin contrast as a function of AC magnetic field strength, for the XY-8 sequence (blue) and for Seq.~B (red). The fit is a sinusoidal oscillation with an exponentially decaying profile~\cite{SM}. Seq.~B shows a steeper slope at zero field, indicating that it is more sensitive than XY-8 to the external field. The interrogation times, $t = 2.16~\mu$s for XY-8 and $t= 6.52~\mu$s for Seq.~B, are independently optimized to achieve maximal sensitivity~\cite{SM}.
{\bf (B)} Extracted absolute sensitivity $\eta$ and volume-normalized sensitivity $\eta_V = \eta\sqrt{V}$ with sensing volume $V = 8.1(9)\times 10^{-3}~\mu$m$^3$ as a function of phase accumulation time $t$ (error bars are given for $\eta$). The pulse spacing $\tau$ is fixed at 25 ns to detect the AC signal oscillating at frequency 11~MHz (see Fig.~3D). A comparison of the two sequences at their respective optimal sensing times reveals that Seq.~B outperforms XY-8 by $\sim$40\%. The solid lines indicate the theoretical sensitivity scaling using the independently estimated sensor characteristics~\cite{SM}.}
\label{fig:fig4}
\end{figure}
Figure~\ref{fig:fig4}A shows the measured contrast $S$ as a function of $B_\text{AC}$ under optimal conditions for Seq.~B and the conventional sequence XY-8, where we choose acquisition parameters for each of the two sequences that optimize their respective absolute sensitivities~\cite{SM}. 
We find that the spin contrast shows faster oscillations and a significantly steeper maximum slope under Seq.~B, indicating that it is more sensitive to the external signal than XY-8. This is due to a combination of optimal state preparation and readout schemes as well as significantly improved coherence times, despite a reduced effective signal strength $|\vec{B}_\text{eff}|$.
In Fig.~\ref{fig:fig4}B, we show the sensitivity scaling with phase accumulation time, for a fixed signal frequency and integration time, finding good agreement with the theoretical prediction~\cite{SM}.
We find that the volume-normalized sensitivity $\eta_V = \eta \sqrt{V}$ of Seq.~B, designed with our optimal sensing approach, reaches more than 40\% improvement in sensitivity over the conventional XY-8 sequence. Moreover, we find that the improvement persists even for sizable (10\%) systematic rotation angle errors, confirming the sequence robustness~\cite{SM}.
With these enhancements, we demonstrate $\eta_V = 8.3(9)$ $\textrm{nT}\cdot\mu\textrm{m}^{3/2}/\sqrt{\textrm{Hz}}$ for Seq.~B, among the best volume-normalized AC sensitivities for solid-state magnetometers measured thus far~\cite{wolf2015subpicotesla} (see Ref.~\cite{SM} for a direct comparison).

Our work establishes a novel approach to quantum sensing by utilizing robust interaction decoupling, and provides the first demonstration of a solid-state ensemble quantum sensor surpassing the interaction limit.
Additionally, our design formalism and generalized effective field picture are also directly applicable to robust DC field sensing.
While our approach already yields a significant improvement in sensitivity, the $T_2$ reached here is still shorter than the depolarization time $T_1\sim100~\mu$s in our dense spin ensemble. This $T_2$ is likely limited by waveform distortions in control pulses and higher-order terms in the average Hamiltonian analysis. The former can be mitigated by waveform engineering~\cite{khaneja2005optimal}, and the latter by sequence symmetrization~\cite{mansfield1971symmetrized} or disorder-reduction via spin-bath engineering~\cite{de2012controlling,bauch2018ultralong}.
Together with new diamond growth techniques~\cite{edmonds2012production}, double-quantum magnetometry~\cite{fang2013high,bauch2018ultralong} and improved photon collection~\cite{wolf2015subpicotesla}, these improvements may push the volume-normalized sensitivity to single-digit picotesla level in a $\mu$m$^3$ volume~\cite{SM}, opening the door to many applications, such as high-sensitivity nanoscale NMR~\cite{aslam2017nanoscale,glenn2018high} and investigations of strongly correlated condensed matter systems~\cite{casola2018probing}.
Beyond applications to diamond magnetometry, the robust sequence design presented in this work can be extended to engineer a broad class of many-body Hamiltonians~\cite{choi2017dynamical,choi2019robust} in a wide variety of quantum hardware platforms, providing a useful tool for quantum information processing~\cite{stajic2013future}, simulation~\cite{zhang2017observation,choi2017observation}, and metrology~\cite{kitagawa1993squeezed,cappellaro2009quantum,choi2018quantum}.

The authors thank W.~W.~Ho, A.~Omran, C.~Ramananthan, T.~Sumarac and E.~Urbach for helpful discussions, A.~Sushko for experimental assistance, and F.~Machado, L.~Martin for critical reading of the manuscript. This work was supported in part by CUA, NSSEFF, ARO MURI, DARPA DRINQS, Moore Foundation GBMF-4306, Samsung Fellowship, Miller Institute for Basic Research in Science, NSF PHY-1506284, Japan Society for the Promotion of Science KAKENHI (No. 26246001), EU (FP7, Horizons 2020, ERC), DFG, SNSF, and BMBF.
H.Z., J.C., S.C., M.D.L. conceived the idea. H.Z., J.C., S.C. performed the  theoretical analysis and pulse sequence design. H.Z., J.C., R.L., A.M.D., H.S.K. performed the experiment and data analysis. J.I., F.J., S.O., H.S. provided the sample. H.Z., J.C., H.S.K. wrote the manuscript, with input from all authors. M.D.L. and H.P. supervised the research.
The authors declare no competing financial
interests.
All data needed to evaluate the conclusions are present in the paper and/or the supplementary materials. Additional data related to this paper may be requested from the authors.
\bibliography{main}

\end{document}


\title{Supplementary Materials for\\ ``Quantum Metrology with Strongly Interacting Spin Systems"}
\date{\today}


\author
{Hengyun Zhou$^{1\ast}$, Joonhee Choi$^{1,2\ast}$, Soonwon Choi$^3$,\\ Renate Landig$^1$, Alexander M. Douglas$^1$, Junichi Isoya$^4$,\\ Fedor Jelezko$^5$, Shinobu Onoda$^6$, Hitoshi Sumiya$^7, $\\ Paola Cappellaro$^8$, Helena S. Knowles$^1$, Hongkun Park$^9$, Mikhail D. Lukin$^{1\dagger}$\\
\normalsize{$^\ast$These authors contributed equally to this work.}
\\
\normalsize{$^\dagger$To whom correspondence should be addressed; E-mail: lukin@physics.harvard.edu}
}

\maketitle
\tableofcontents

\section{Experimental details}
\label{sec:experiment}

\subsection{Diamond sample}
\begin{figure*}[h]
\begin{center}
\includegraphics[width=1\textwidth]{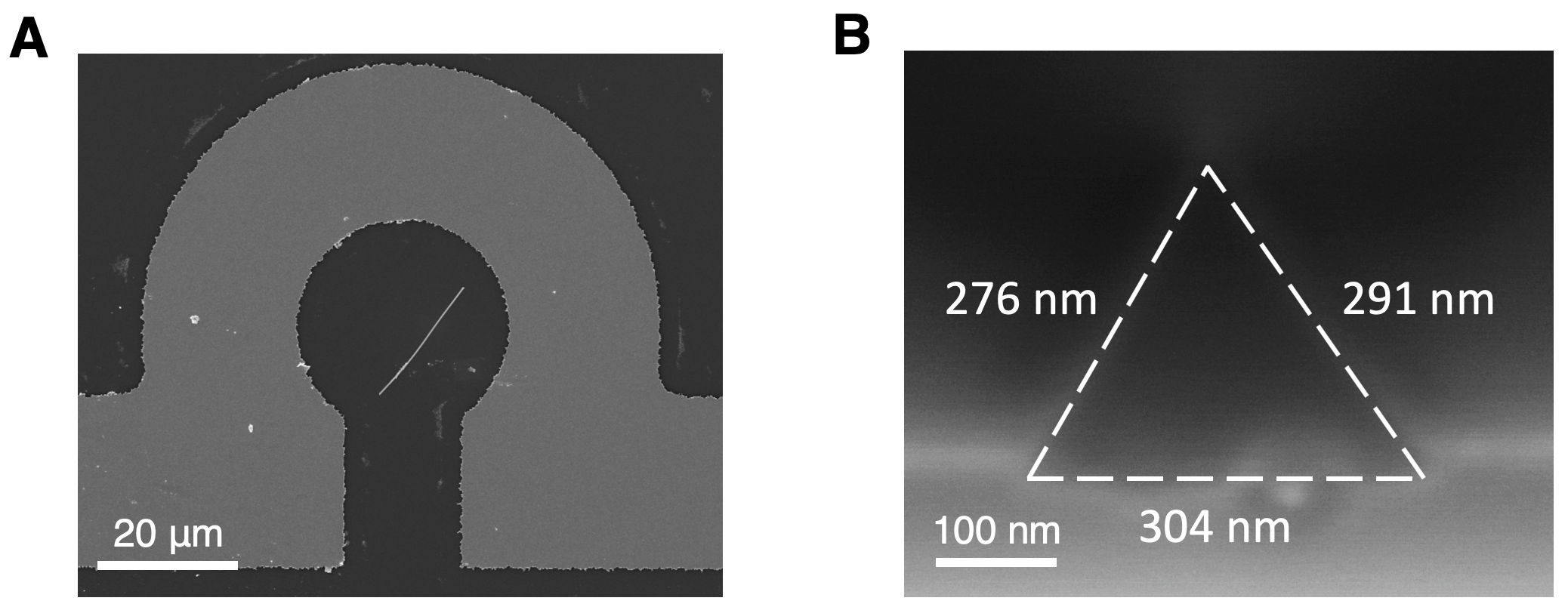}
\end{center}
\caption{Scanning electron microscope images of the diamond nanobeam used in our experiments. (\textbf{A}) A typical nanobeam sample (thin line at the center) placed near the center of an $\Omega$-shaped microwave delivery line (bright region) fabricated on a glass substrate (dark regions), (\textbf{B}) SEM cross-section of the nanobeam. The white boundary (dashed line) indicates the effective area used for sensor volume estimation, and the measured lengths of each edge are labeled, with an error of $\pm10$ nm on each value.}
\label{fig:nanobeam}
\end{figure*}

The sample used in this work is a type-Ib, high-pressure high-temperature (HPHT) diamond with a total negatively-charged nitrogen-vacancy (NV$^-$) concentration of $\sim$15 ppm (see {\it Characterization of NV$^-$ Disorder and Interaction Strength} below). To achieve such a high density, the sample was irradiated with high-energy electron beams (2 MeV) at a flux of $1.3 \times 10^{13}~\text{cm}^{-2} \text{s}^{-1}$ and simultaneously in-situ annealed at $700-800^{\circ}\text{C}$ for a total of 285 hours to reach a total fluence of $1.4 \times 10^{19}~\text{cm}^{-2}$. This leads to a high conversion efficiency ($>10\%$) of the initial nitrogen into NV centers. The average distance between NV centers and the corresponding NV-NV interaction strength, taking into account all NV lattice orientations, are estimated to be $\sim$7~nm and $\sim(2\pi)$~140~kHz, respectively~\cite{kucsko2018critical}. The unpaired leftover nitrogen (P1 centers) act as paramagnetic defects, creating a local magnetic field to nearby NV centers. Due to the presence of such impurities, as well as $^{13}$C nuclear spins, local charges, and strain in the sample, the linewidth of the NV resonance is inhomogeneously broadened to $(2\pi)~4.0$ MHz (Gaussian standard deviation), corresponding to on-site disorder terms. Due to strong inhomogeneous broadening, the hyperfine interaction of strength $(2\pi)~2.2$~MHz between the NV electronic spin and the $^{14}$N host nuclear spin is not resolved in the measured optically-detected magnetic resonance spectrum (see Fig.~S3A). To improve control homogeneity, we fabricated a $20~\mu$m-long nanobeam structure from bulk diamond via Faraday cage angled etching~\cite{burek2012free}. As shown in Fig.~\ref{fig:nanobeam}, the nanobeam has a triangular cross section, with dimensions labeled in the figure.

Adopting a definition similar to the way mode volumes are defined, we define the effective probing area $A$ to be the integrated power within the 2D Airy pattern divided by the maximum intensity at the center of the beam. Using this definition, we have
\begin{align}
A = \pi r^2=\int_0^\infty ds \; 2\pi s \left(\frac{2 J_1(2\pi s (\text{NA})/\lambda)}{2\pi s (\text{NA})/\lambda}\right)^2=\frac{\lambda^2}{(\text{NA})^2\pi},
\end{align}
where $r$ is the effective probing radius, $\textrm{NA}=1.3$ is the numerical aperture of the objective, $\lambda=532$ nm is the excitation wavelength, and $J_1$ is the Bessel function of the first kind of order one. Plugging in these values gives $r=\lambda/(\pi\cdot \text{NA})\approx 130$ nm.

Denoting the bottom edge length as $c$, the height as $h$, we estimate the probing volume by integrating over the intersection of a cylinder of radius $r$ with the profile of the nanobeam. Since $2r<c$, the bottom part will be an integration over a circular profile, while the upper parts will be cut off due to the edges of the triangular nanobeam. This gives the following integration for the volume:
\begin{align}
V&=\pi r^2 \frac{h(c-2r)}{c}+\int_0^{\pi/2}d\theta\frac{2r}{c}h\cos\theta(2r^2\cos\theta\sin\theta+2r^2\theta)\approx 8.1\times 10^{-3}~\mu\text{m}^3.
\end{align}

Assuming a 10 nm uncertainty on each of the parameters of the nanobeam, we can perform error propagation to arrive at a probing volume of $V\approx (8.1\pm 0.9)\times 10^{-3}~\mu\text{m}^3$.

\subsection{Optical and microwave setup}
The optical setup consists of a room temperature home-built confocal microscope with an oil-immersion objective (Nikon 100x, NA = 1.3). The sample is mounted on a piezoelectric stage in the focal plane of the microscope. A green laser (532 nm) is used to excite the ensemble of NV centers inside the confocal volume. To switch the excitation laser on and off, an acousto-optic modulator (Gooch \& Housego, 3200-125) is installed in a double-pass configuration. NV centers emit fluorescence into the phonon sideband (630-800 nm), which is isolated from the excitation laser by a dichroic mirror. After passing through a 650 nm long-pass filter, the emitted photons are fiber-coupled into a multi-pixel photon counter (MPPC, Hamamatsu C14452-1550GA), which has a large dynamic range to accomodate our high photon count rates, and a quantum efficiency of $\sim$25$\%$ in the wavelength range of interest. For the sensing measurements, we optimize the green laser power to maximize the signal-to-noise ratio (SNR) (see the section IV. B. {\it sensitivity optimization}).

A high sampling rate arbitrary waveform generator (Tektronix, model AWG7122C) defines waveforms of the pulses with a temporal resolution of 83 ps (=12 Giga-samples per second). Using two separate channels of the AWG, we synthesize $\pi/2$ and $\pi$ pulses at a calibrated Rabi frequency of $\Omega=25$ MHz ($\pi/2$ pulse length of 10 ns) and generate a continuous-wave target sensing signal oscillating at an arbitrary frequency. The synthesized pulses and the sensing signal are separately amplified by microwave amplifiers (Mini-Circuits, ZHL-16W-43-S+ and LZY-22+, respectively). To reduce the nonlinearity of the microwave amplifier, a DC block (Picosecond, 5501a) and low-pass filter (Mini-Circuits, VLF-3000+) are added. The amplified microwaves are combined together by a diplexer (Microwave Circuits, D3005001) before being delivered to a coplanar waveguide where the NV-ensemble is located, with an estimated $1\%$ inhomogeneity in Rabi frequency across the probing volume~\cite{choi2017observation}. The AC magnetic field target sensing signal will have the same amount of inhomogeneity, which will be neglected in the following, as the broadening associated with it is minimal. Attenuators and a circulator (Pasternack, PE83CR1004) are also inserted at various places along the microwave delivery line to isolate unwanted reflections at the interfaces between different cables and devices.

\subsection{Experimental parameters and measurement sequences}
The diamond sample contains four subgroups of NV centers, each oriented along one of the four different crystallographic axes of the crystal. Each NV center has a magnetic-field-sensitive spin triplet in its electronic ground state, $\ket{m_s=0, \pm1}$, which can be coherently manipulated via microwaves. The combination of a permanent magnet and electromagnetic coils is used to produce a static magnetic field that adjusts relative energy spacings between the spin states via Zeeman shifts. In the experiments presented in the main text, the orientation of the static field is set parallel to the crystallographic axis of a single NV group with a magnitude of 260~gauss. Due to their differing dipole orientations, the excitation efficiency for different NV groups will vary slightly, resulting in different spin contrasts; consequently, we use the NV group experimentally-determined to have the largest contrast and align the external magnetic field to this NV group direction. In such a setting, the transition frequencies of the NV centers from $\ket{0}$ to $\ket{-1}$ and $\ket{0}$ to $\ket{+1}$ correspond to $\omega_1= 2.137$~GHz and $\omega_2= 3.602$~GHz, respectively. Illumination by a green laser initializes the NV spin state into $\ket{0}$ via a state-selective intersystem crossing~\cite{goldman2015phonon}. A moderate laser power of around 50~$\mu$W is chosen to suppress fluorescence fluctuations associated with charge state instabilities in the dense NV ensemble while maintaining good fluorescence count rates~\cite{choi2017depolarization}. 

\begin{figure*}[h]
\begin{center}
\includegraphics[width=1\textwidth]{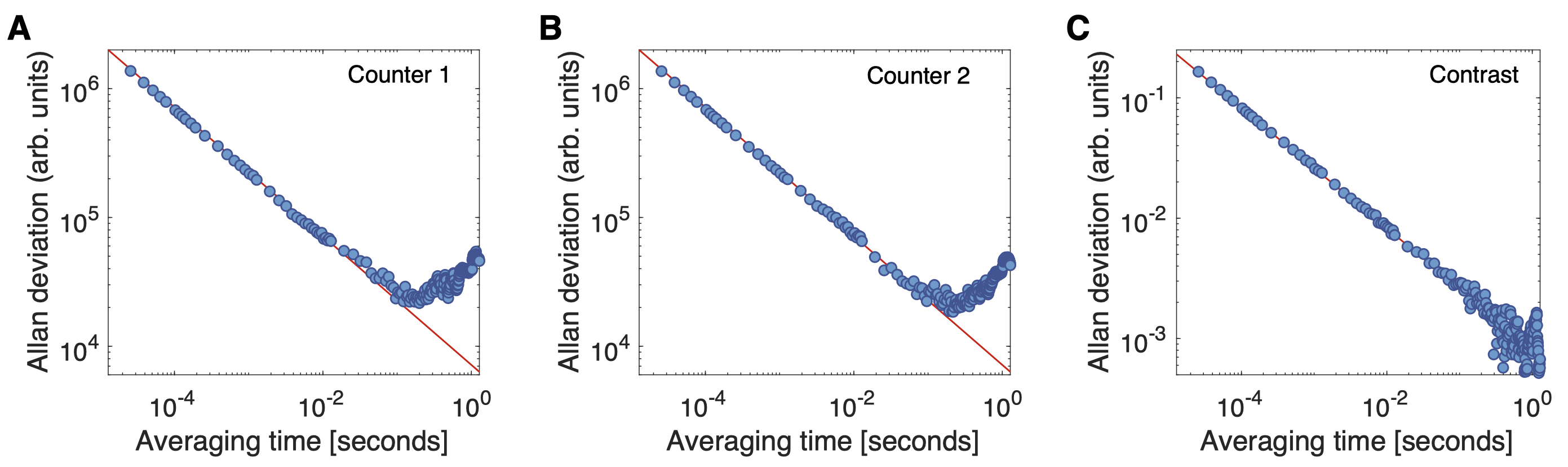}
\end{center}
\caption{Allan deviations for (\textbf{A,B}) individual photon counters $c_{1,2}$ and (\textbf{C}) contrast $S$. The red solid line denotes a $1/\sqrt{\tau}$ scaling fit where $\tau$ is the averaging time. As seen in (\textbf{C}), the contrast $S = 2\qty(c_1-c_2)/\qty(c_1+c_2)$ effectively rejects common-mode noise between the consecutive counters, giving rise to an improved stability of the sensor.}
\label{fig:allan}
\end{figure*}

In each measurement, illumination by a green laser initializes the NV spin state into $\ket{0}$ via a state-selective intersystem crossing~\cite{goldman2015phonon}. After initialization, a series of microwave pulses resonant with the $\ket{0} \leftrightarrow \ket{-1}$ transition are applied to create coherent superpositions of the two states and control the effective two-level dynamics. As illustrated in Fig.~2A of the main text, each pulse sequence is repeated twice, with and without a final $\pi$ pulse acting on the $\ket{0} \leftrightarrow \ket{-1}$ transition. At the end of each sequence, the photons emitted from NV centers are detected for a short duration ($t_\text{read} = 0.5~\mu$s, chosen to maximize the SNR at our operating laser power) to read out their spin states. If we define $c_1$ and $c_2$ as the two counters recording the photon numbers with and without the $\pi$ pulse, respectively, then the contrast $S$ is defined as $S = 2(c_1 - c_2)/(c_1 + c_2)$. Physically, $S$ is linearly proportional to the population difference between $\ket{-1}$ and $\ket{0}$. Such a differential measurement is beneficial for suppressing common noise sources between the two counters, improving the SNR. To confirm this, we experimentally characterize the Allan deviation for the individual counters $c_{1,2}$ as well as the contrast $S$. The Allan deviation is known to estimate the long-term stability of a sensor~\cite{degen2017quantum}. As shown in Fig.~\ref{fig:allan}, the individual counters start to deviate from the expected $1/\sqrt{\tau}$ scaling when the averaging time $\tau$ is longer than $\sim0.1$ seconds, indicating that further averaging does not lead to an improvement in SNR. However, the Allan deviation of the contrast $S$ continues to follow the desired scaling, suggesting that the contrast measurement cancels out common-mode noise present in adjacent counters.

For the spin-echo measurement, we vary the free evolution time and monitor the contrast as a function of the total evolution time. For the other dynamical decoupling measurements, we fix the pulse separation $\tau=25$~ns and monitor the contrast as a function of the number of repetitions of the dynamical decoupling block (equivalently, the total evolution time). We fit the decay profile with a stretched exponential $S(t)=S_0\exp[-\qty(t/T_2)^\alpha]$, and plot the normalized contrast $S(t)/S_0$ as the coherence in Fig.~2C of the main text. Note that due to the different types of sweeping ($\tau$ vs. Floquet repetition number) performed for the spin echo and other dynamical decoupling sequences, the stretched exponential power $\alpha$ is rather different. For these measurements, we choose a low experimental duty cycle and a green laser power of 50~$\mu$W. More specifically, we use a readout green pulse of 5~$\mu$s with the photon counter duration being 0.5~$\mu$s, followed by a 100~$\mu$s wait time for full equilibration of the charge dynamics \cite{choi2017depolarization}, and then a 20~$\mu$s green pulse to repolarize the NV centers.

For the sensing measurements, we use a fixed pulse separation $\tau=25$~ns, and choose the frequency of the AC sensing signal field to be at the most sensitive resonance, with its frequency calculated by taking into account the frequency-domain modulation function with experimental parameters for the pulse durations and spacings (see Eq.~(3) of the main text as well as Sec.~\ref{sec:sequencedesign} and Ref.~\onlinecite{otherpaper}). In addition, we experimentally verify that the chosen frequencies give rise to maximal signal contrast. In order to maximize the SNR, we fully optimize all measurement parameters and minimize the experimental overhead times. For this purpose, we utilize a single green pulse (50 $\mu$W) for both initialization and readout, and optimize its duration to be 4~$\mu$s, balancing the time overhead and imperfect repolarization effects of the NVs (see Sec.~\ref{sec:sensopt} for details). To guarantee a reset of the NV sensor, we insert a depolarizing $\pi/2$ pulse right after readout to eliminate any remnant polarization of the NV centers, preventing any correlations between neighboring measurements. The fixed duty cycle in these measurements ensures that charge dynamics does not affect the measurement results.

For each sensing sequence, we first fix the AC sensing signal frequency to the expected resonance frequency and optimize the relative phase between the sensing signal and the sequence to achieve maximal contrast (see Sec.~\ref{sec:relphase}). We then sweep the signal frequency to experimentally confirm the location of the resonance (see for instance Fig.~3D of the main text), and perform another iteration of phase optimization at the newly identified resonance frequency. Note that during this frequency sweep, the phase is carefully adjusted to prevent any additional artifacts arising from phase differences at different frequencies. Having confirmed the resonance frequency and optimal signal phase, we sweep the AC signal amplitude and measure the contrast oscillations as a function of AC signal magnetic field strength (see Sec.~\ref{sec:fieldcalibration} for the calibration procedure of the field strength). Taking the maximum slope of the oscillations (typically achieved at zero field for sine magnetometry) and evaluating the fluctuations in each data point, as well as considering the measurement duration, we can extract the sensitivity, see Sec.~\ref{sec:sensitivity} for more details. The entire procedure is then repeated for a number of sequence repetition cycles to obtain the optimal sensitivity for a given sequence, balancing the effects of longer phase accumulation times and increased decoherence.

\subsection{Characterization of NV$^-$ Disorder and Interaction Strength}
In order to characterize the on-site disorder and interaction strength of our NV ensemble, we compare the experimentally-measured decay profile of a Ramsey sequence and XY-8 sequence to numerical simulations. We consider a system of $N$ spin-1/2 particles with on-site disorder and dipolar interactions, with Hamiltonian
\begin{align}
H = H_s + H_\Omega(t),
\end{align}
where the system Hamiltonian $H_s$ and the time-dependent control field $H_\Omega(t)$ can be described as
\begin{align}
	H_s &= \sum_i h_i S_i^z - \sum_{ij} \frac{J_0}{r_{ij}^3} (1-3\cos^2\theta_{ij}) \left( S_i^x S_j^x +  S_i^y S_j^y - S_i^z S_j^z  \right), \\
	H_\Omega(t) &= \sum_i \Omega^x (t) S_i^x  + \Omega^y (t) S_i^y.
\end{align}
Here, $S_i^\mu$ with $\mu \in \{x,y,z\}$ are spin-1/2 operators for spin $i$, $h_i$ is an on-site disorder Zeeman potential, $\Omega^{x/y}$ are Rabi frequencies of the microwave driving along $\hat{x}/\hat{y}$-axes, $J_0 = (2\pi)~52$\,MHz$\cdot$nm$^3$ is the coefficient of dipolar interactions between two electronic spins, and $r_{ij}$, $\theta_{ij}$  are the distance and relative orientation between the two spins at sites $i$ and $j$.

For simulations, we numerically integrate the time evolution under the Hamiltonian for various pulse-sequences. $\pi$- or $\pi/2$-pulses along either $\hat{x}$- or $\hat{y}$-axes are implemented by setting the corresponding Rabi frequency $\Omega^x = \Omega^y = (2\pi)\times 41.7$\,MHz and evolving the system for an appropriate time duration $t_p$, thus incorporating the effects of both interactions and disorder during these pulses. The free evolution time is chosen to be $\tau=20$ ns, in accordance with this set of experiments. We provide exact numerical results for $N = 16$ particles and assume that the $N$ spins are randomly positioned within a 3D box of dimension $L\times L \times L$ ($L \equiv n^{-1/3}$) with periodic boundary conditions, where $n$ is the spin density, and require that no pair is closer than a cut-off distance $r_\text{cut} = 1.4$\;nm. Physically, this cut-off distance arises because any pair of NV centers much closer than $r_\text{cut}$ will be severely affected by direct charge tunneling, which in turn significantly modifies their energy level structure~\cite{chou2018first}. Due to the finite system size, the simulations exhibit a residual polarization at long times, which we subtract out for comparison to experiments.

\begin{figure*}[h]
\begin{center}
\includegraphics[width=0.95\textwidth]{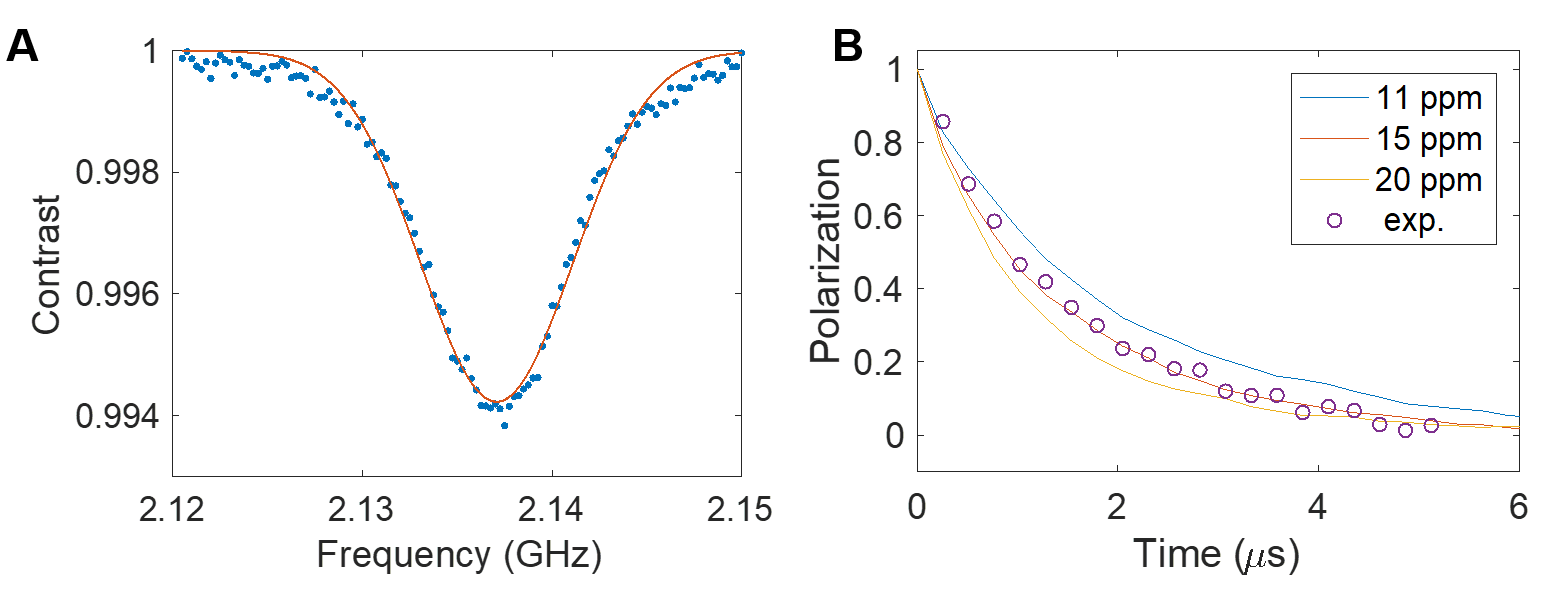}
\end{center}
\caption{{\bf (A)} Measured ESR for the NV-spin ensemble, characterizing the on-site disorder distribution, with an inverted Gaussian fit (red trace) and {\bf (B)} Comparison of experimental decay profiles under the XY-8 sequence to simulation results for different total spin densities. See text for more details.}
\label{fig:density}
\end{figure*}

As shown in Fig.~\ref{fig:density}A, we first measure the random on-site disorder via continuous-wave electron spin resonance, which is well-described by a Gaussian distribution with standard deviation $\sigma_W = (2\pi)~4.0\pm0.1$~MHz. Using this disorder strength, we then turn to an XY-8 sequence to extract the spin density by comparing simulated decay curves to the experimental observations (Fig.~S3B). We expect the decoherence rate under the XY-8 sequence to be dominated by spin-spin interactions as the sequence is designed to efficiently decouple the static disorder, which we have also directly verified below, where the coherence time under an XY-8 sequence scales proportionally to the number of resonant NVs. By varying spin densities in the simulations, we find that a total spin density of around 15 ppm best reproduces the experimental decay profile (Fig.~\ref{fig:density}B). We note however that this estimation is only a rough approximation to the actual spin density, as the simulations in such a small system may not fully reflect the effect of long-range interactions, especially for spin ensembles in three dimensions. Our previous density extraction based on the spin-echo sequence resulted in a higher estimated spin density of $\sim$45 ppm~\cite{kucsko2018critical}, indicating that the numerically-extracted density values depend on the details of the modeling of finite-size spin systems as well as the methods employed to quantify the spin density.

\begin{figure*}[h]
\begin{center}
\includegraphics[width=0.95\textwidth]{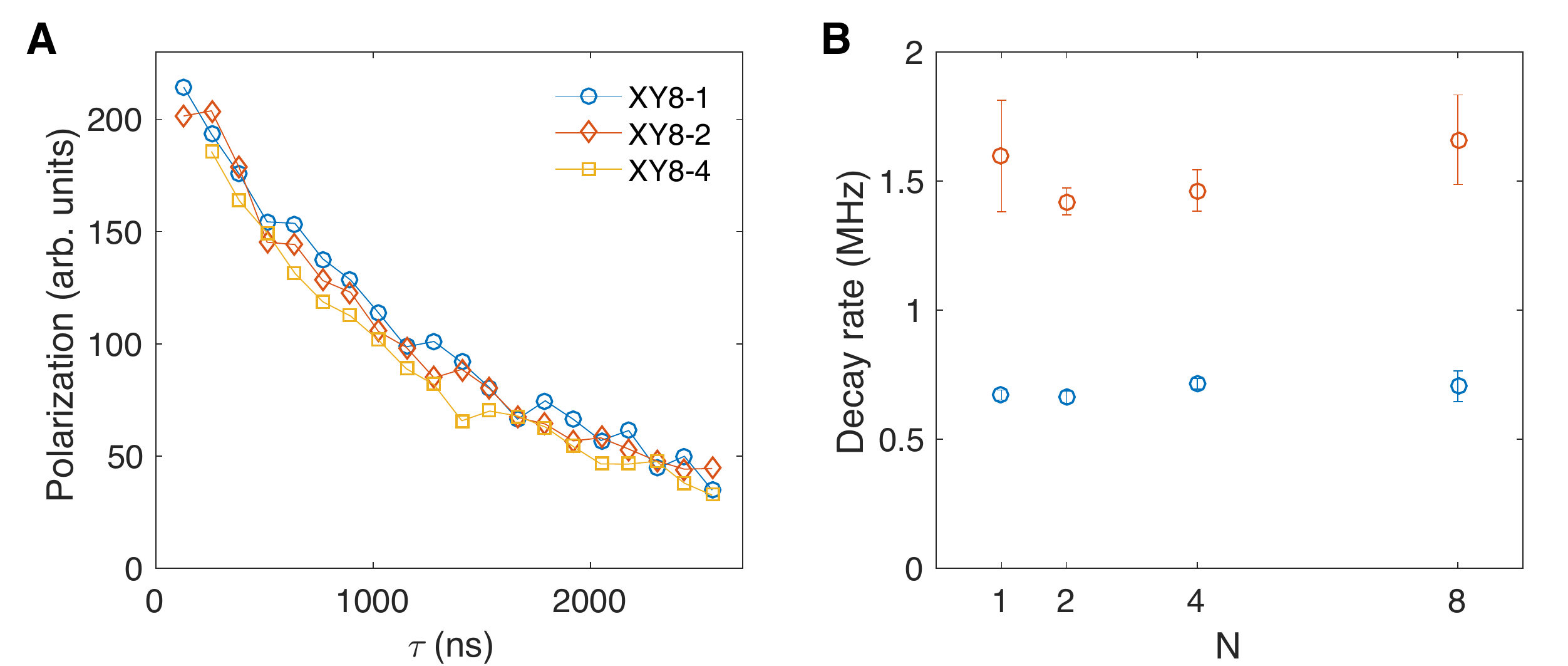}
\end{center}
\caption{{\bf (A)} Coherence decay as a function of total time, for different numbers of repeated XY-8 blocks. Different numbers of pulses give rise to the same decay curve, indicating that the coherence decay is not a result of spin-bath contributions. {\bf (B)} Decay rates as a function of the number of repeated XY-8 blocks, for a single group of NVs (blue) and two groups of NVs tuned onto resonance (red). The decay rate for two groups of NVs on resonance is approximately twice that of a single group, consistent with the factor of two difference in density.}
\label{fig:groupxy}
\end{figure*}

To further verify that the observed XY-8 coherence time is interaction-limited, we examine the coherence time as a function of the number of repeated XY-8 blocks, as well as for different NV densities achieved by tuning multiple groups of NVs onto resonance. As shown in Fig.~\ref{fig:groupxy}A, different numbers of repeated XY-8 blocks give rise to the same decay curve; since a larger number of repeated XY-8 blocks implies a smaller pulse spacing and more efficient decoupling from spin-bath contributions, the fact that the coherence decay is independent of the number of pulses indicates that the coherence time is not spin-bath limited. Furthermore, we find that by tuning two groups of NV centers with different lattice orientations onto resonance, thus doubling the spin density, the decay rate of the XY-8 sequence is doubled (Fig.~\ref{fig:groupxy}B), demonstrating that the coherence time is indeed interaction-limited. Since the magnetic field sensitivity is inversely-proportional to the coherence time, this also directly implies that the XY-8 sensing sequence sensitivity is interaction-limited, and by using our newly-developed formalism we are surpassing the sensitivity interaction limit.

\section{Pulse sequence design and performance analysis}
\subsection{Summary of design formalism}
\label{sec:sequencedesign}
In this section, we summarize the key ingredients of our pulse sequence design framework. Theoretical extensions to the framework, as well as its use in other applications and different system Hamiltonians, can be found in Ref.~\onlinecite{otherpaper}. The key idea of our approach is to adopt a description of the pulse sequence in terms of how the $S^z$ spin operator in the interaction picture is transformed during the free evolution periods~\cite{mansfield1971symmetrized}, also known as the toggling frame picture. Using this pulse sequence description, we generalize previous techniques developed in the nuclear magnetic resonance (NMR) community to generic spin systems with general types of interactions, with very different design requirements, including our disorder-dominated dipolar spin ensembles. Crucially, we find that this enables a simple algebraic description of both the conditions for on-site disorder and interactions to be decoupled during the free evolution periods, and for the systematic suppression of many kinds of finite pulse duration effects, including the action of on-site disorder and residual interactions during the pulse, as well as rotation angle errors. Moreover, this description is also naturally suited to analyze the sensing properties of the sequence. Utilizing these simple algebraic conditions for decoupling and sensing, we are able to design different pulse sequences with varying decoupling performances, with Seq.~A and Seq.~B in the main text as specific examples. We note that existing strategies to construct robust decoupling pulse sequences, such as the use of composite pulses~\cite{levitt1986composite}, are largely incorporated into our formalism. For example, a large class of such composite pulses based on the combination of $\pi/2$-rotations around $\hat{x}$, $\hat{y}$ axes are already included (see section II. B. {\it detailed analysis of pulse sequences}). However, for more complicated versions where a large number of sequential rotations with fine-tuned angles and axes are used, it is expected that the extended length of composite pulses will significantly limit the interaction-decoupling efficiency, as they are typically optimized for single-spin rotations in non-interacting systems, and thus we expect sequences designed with our formalism to perform better.

We start by describing the pulse sequence representation, and the set of algebraic rules imposed on the representation that guarantee leading-order fault-tolerant dynamical decoupling. A given pulse sequence, consisting of $n$ $\pi/2$ pulses with finite pulse duration $t_p$ as building blocks (a $\pi$ pulse is treated as two $\pi/2$ pulses with zero time separation in between), $\{P_k\}$, $k=1,2,\cdots,n$, is represented by a frame-duration vector $\boldsymbol{\tau}=[\tau_k]$ and a 3-by-$n$ frame matrix ${\bf F} =  [F_{\mu, k}] = [\vec{F}_x; \vec{F}_y; \vec{F}_z]$. The elements $\tau_k$ of the frame-duration vector indicate the free evolution duration preceding pulse $P_k$. The elements $F_{\mu,k}$ of the frame matrix characterize the form of the transformed Pauli spin operator $\tilde{S}^z_k$ in the free evolution duration preceding pulse $P_k$. More precisely, we define
\begin{align}
\tilde{S}^z_k &= (P_{k-1} \cdots P_1)^\dagger S^z (P_{k-1} \cdots P_1)=\sum_\mu F_{\mu,k} S^\mu,
\end{align}
from which the frame matrix can also be easily obtained by inverting the relation:
\begin{align}
F_{\mu,k} = 2\Tr[S^\mu \tilde{S}^z_k] \quad \text{for $\mu = x,y,z.$}
\label{eq:modtime}
\end{align}

Based on this representation, we can easily formulate the conditions for disorder and interactions to be decoupled, and for all  types of leading-order finite pulse imperfections to be suppressed~\cite{otherpaper}. The conditions are summarized as follows:
\setlist[enumerate]{wide=0pt, leftmargin=15pt, labelwidth=15pt, align=left}
\renewcommand{\theenumi}{\bf{\quad\quad Rule \arabic{enumi}}}
\begin{enumerate}
\item \textbf{Decoupling of on-site disorder and anti-symmetric spin-exchange interactions:}
\begin{equation}
\sum_{k=1}^n F_{\mu, k} \qty(\tau_k+\frac{4}{\pi}t_p) =0 \quad \text{for every $\mu=x,y,z.$}
\label{eq:rule1}
\end{equation}
\item \textbf{Symmetrization of Ising and symmetric spin-exchange interactions:}
\begin{equation}
	\sum_{k=1}^n |F_{\mu,k}|(\tau_k+t_p) \quad \text{is the same for all $\mu = x,y,z.$}
    \label{eq:rule2}
\end{equation}
\item \textbf{Decoupling of interaction cross-terms (parity cancellation):} 
\begin{equation}
 \sum_{k=1}^n (F_{\mu,k} F_{\nu,k+1} + F_{\nu,k}F_{\mu,k+1})=0 \quad \text{for all pairs $(\mu,\nu).$}
 \label{eq:rule3}
\end{equation}
\item \textbf{Suppression of rotation angle error (chirality cancellation):} 
\begin{equation}
\sum_{k=1}^n \vec{F}_{k+1} \times \vec{F}_{k} = \vec{0} \quad \text{where $\vec{F}_k =\sum_\mu F_{\mu,k} \hat{e}_\mu$, with $\hat{e}_\mu$ being the $\mu$-axis unit vector.}
\label{eq:rule4}
\end{equation}
\end{enumerate}

\renewcommand{\theenumi}{\arabic{enumi}}

While the full theoretical derivation of the above design rules can be found in Ref.~\onlinecite{otherpaper}, here we sketch the main idea of it. As discussed in the main text, a given periodic pulse sequence \{$P_k; k = 1,2,.., n$\} with free evolution time before each pulse given by \{$\tau_k; k = 1,2,.., n$\} defines a unitary operator $\mathcal{U}(T) = P_k e^{-iH_s \tau_k} \cdots P_1 e^{-iH_s \tau_1}$ over one cycle. Defining the toggling-frame Hamiltonians $\tilde{H}_k=(P_{k-1} \cdots P_1)^\dagger H_s (P_{k-1} \cdots P_1)$, the unitary operator can be rewritten as
\begin{align}
\mathcal{U}(T)=e^{-i\tilde{H}_k\tau_k}\cdots e^{-i\tilde{H}_2\tau_2}e^{-i\tilde{H}_1\tau_1} \approx e^{-iH_\text{avg}T},
\end{align}
where the leading-order average Hamiltonian is $H_{\text{avg}} = \frac{1}{T}\sum_{k=1}^n \tilde{H}_k \tau_k$ and $T$ is the Floquet period. The leading-order description is a good approximation to the dynamics when the driving frequency $1/T$ is much faster than the local energy scales of the system Hamiltonian $H_s$. To further increase accuracy, higher-order corrections to this expression can be systematically accounted for using the Floquet-Magnus expansion~\cite{magnus1954on}.

The internal system Hamiltonian of NV ensembles includes disorder and interaction terms, given in Eq.~(S4), and can be rewritten as
\begin{align}
H_s=\sum_{i} h_{i} S_{i}^{z}+\sum_{i j} \frac{J_{i j}}{r_{i j}^{3}}\left(\vec{S}_{i}\cdot\vec{S}_{j}-2S_{i}^{z} S_{j}^{z}\right),
\end{align}
where $h_i$ is the on-site disorder field for the spin at site $i$ and $J_{ij}/r_{ij}^3$ is the dipolar interaction strength between two spins at sites $i$ and $j$. As the Heisenberg component of the interaction Hamiltonian, $\vec{S}_{i}\cdot\vec{S}_{j}$, is invariant under global rotations, the $k$-th toggling-frame Hamiltonian $\tilde{H}_k$ can be uniquely specified by how the $S^z$ spin operator is transformed after the preceding $k-1$ pulses:
\begin{align}
\tilde{H}_k=\sum_{i} h_{i} (\tilde{S}_k^{z})_i+\sum_{i j} \frac{J_{i j}}{r_{i j}^{3}}\left(\vec{S}_{i}\cdot\vec{S}_{j}-2(\tilde{S}_k^{z})_i (\tilde{S}_k^{z})_j\right).
\end{align}
As discussed above, $\tilde{S}^z_k =\sum_\mu F_{\mu,k} S^\mu$. Therefore, the matrix $F_{\mu,k}$ allows us to directly write down the average Hamiltonian, $H_{\text{avg}} = \frac{1}{T}\sum_{k=1}^n \tilde{H}_k \tau_k$ describing spin evolution. This method is in fact generally applicable to any Hamiltonian under a strong quantizing field, with up to three-body interactions~\cite{otherpaper}.

With this analysis, we can now understand the intuition behind the above rules for leading-order fault-tolerant dynamical decoupling of disorder and interactions. 

First, the cancellation of disorder [Rule 1] can be understood as the requirement that a spin echo is performed along each axis direction, such that a precession around a positive disorder axis (e.g. $+S^z$) is compensated by a negative precession ($-S_z$). This suggests that there should be an equal amount of free evolution time along the positive and negative directions for each axis. In addition to the contributions from free evolution periods, the disorder will also be acting during the finite pulse duration. Crucially, since the spin operator continuously rotates during the $\pi/2$ pulse (e.g. $S^z \cos\theta + S^x \sin\theta$ when rotating from $S^z (\theta = 0)$ to $S^x (\theta = \pi/2)$), the disorder acting during the finite pulse duration results in a contribution that is proportional to the disorder Hamiltonian immediately preceding and following the pulse, with an additional numerical prefactor $2/\pi$ after integrating over the pulse. Thus, the disorder Hamiltonian of the pulse $P_k$ can be written as $\bar{H}_{P_k} = \frac{2}{\pi} (\tilde{H}^\text{dis}_{k-1} + \tilde{H}^\text{dis}_{k})$. This effectively extends the disorder contribution from the $k$-th free evolution period $\tau_k$ to a duration of $\tau_k + 4t_p/\pi$. By treating $\pi$ pulses or composite $\pi/2$ pulses as a combination of several $\pi/2$ pulses with zero time separation in between, and specifying the intermediate frame direction after each $\pi/2$ pulse, we ensure that the effects of finite pulse duration during all pulses are treated appropriately.

Similarly, the suppression of interaction effects [Rule 2] in our Hamiltonian can be easily understood from a generalization of the WAHUHA sequence~\cite{waugh1968approach}. Here, by rotating the $S^z$ spin operator to spend equal time along the $\hat{x}$, $\hat{y}$ and $\hat{z}$ directions, the interactions are symmetrized into the Heisenberg form, which preserves the coherence of polarized initial states that are typically realized in ensemble experiments. The effects of finite pulse duration can again be analyzed by considering the continuous rotation of the spin operator; however, as the interaction involves two operators, there will be both a contribution that extends the effective duration of each free evolution period, and an interaction cross-term involving the axis directions before and after the pulse. The interaction cross-term cancellation can be easily phrased as a parity product condition between neighboring frame directions [Rule 3].

Furthermore, robustness against different control imperfections can also be readily incorporated with our representation. Here, we focus on the dominant source of control error in many experimental systems, namely rotation angle errors that may arise due to an imperfection calibration of microwave power or control field spatial inhomogeneities, although other types of imperfections can also be directly incorporated. Here, the intuition is that the average Hamiltonian contribution corresponding to a rotation angle error can be easily analyzed \textit{in the toggling frame}, where a systematic over-/under-rotation along the $+\hat{\mu}$-axis (positive chirality) can be compensated by another rotation along the $-\hat{\mu}$-axis (negative chirality) in the toggling frame. This chirality can be conveniently described mathematically by the cross product between the neighboring frame directions, allowing a simple algebraic rule for the cancellation of rotation angle errors [Rule 4].

Combining these, we have a set of succinct algebraic conditions to describe leading-order fault-tolerant dynamical decoupling, which can enable a significant extension of spin coherence times. However, to perform effective sensing, it is also crucial to maintain high sensitivity to an AC external signal. We now show how the representation we introduced above also readily enables the analysis of sensing properties of a given sequence, allowing us to design pulse sequences that approach the optimal sensitivity to an external field under the constraints of interaction-decoupling.

First, we derive the average Hamiltonian contribution of the target AC sensing signal. For an AC magnetic field signal causing a time-dependent sinusoidal line shift, the Hamiltonian is given by
\begin{align}
H_{\text{AC}}(t)=\gamma_\text{NV} B_{\text{AC}}\Re\qty[e^{-i (2\pi f t-\phi)}] \sum_i S_i^z,
\end{align}
where $\Re$ denotes taking the real part, $B_\text{AC}$, $f$, $\phi$ are the amplitude, frequency and phase of the AC magnetic field. Note that under the secular approximation in the presence of a strong quantizing field, only the projection of $B_\text{AC}$ onto the quantization axis will be relevant. To capture the resonance characteristics of a given pulse sequence, we define the frequency-domain modulation functions
\begin{align}
\tilde{F}_\mu(f) =|\tilde{F}_\mu(f)| e^{-i\tilde{\phi}_\mu(f)} = \frac{1}{NT} \int_0^{NT} e^{-i2\pi f t'}F_{\mu}(t')dt',
\end{align}
where $F_{\mu}(t)$ is the time-domain modulation function, as defined in Eq.~(\ref{eq:modtime}) but accounting for the finite pulse duration effects. $N$ is the sequence repetition number, $T$ is the duration of the Floquet period, and $\tilde{\phi}_\mu(f)$ are spectral phases capturing the relative phase differences between different axes. The average Hamiltonian contribution corresponding to this sensing field can be written as
\begin{align}
H_\text{avg,\text{AC}} = \gamma_{\text{NV}} \vec{B}_\text{eff}(f) \cdot \sum_i \vec{S}_i,
\end{align}
where the {\it frequency-dependent} vectorial effective sensing field $\vec{B}_\text{eff}$ is expressed as
\begin{align}
\vec{B}_\text{eff}(f)= B_\text{AC} \sum_{\mu=x,y,z} \qty|\tilde{F}_\mu(f)| \Re[e^{-i(\tilde{\phi}_\mu(f) - \phi)}] \vec{e}_\mu.
\end{align}
This implies that under the influence of the target AC sensing field, the spin will precess around an effective magnetic field $\vec{B}_\text{eff}$ that is determined by the vector frame modulation functions of the sensing sequence. We note that this effective sensing field description can also be directly applied to the analysis of DC sensing sequences. In this vectorial sensing scheme, the total spectral response $|F_t|$ is evaluated as
\begin{align}
\qty|\tilde{F}_t(f)|=\frac{|\vec{B}_\text{eff}(f)|}{B_\text{AC}}=\sqrt{\frac{1}{2}\qty[\sum_\mu \qty|\tilde{F}_\mu(f)|^2 + \sum_\mu \Re \qty[\tilde{F}_\mu(f)^2]]}.
\end{align}
Thus, to optimize sensitivity, we require that the frequency-domain modulation functions $|\tilde{F}_\mu(f)|$ along each of the three axis directions have resonances aligned at the same frequency $f$, and that the phase of the resonance $\tilde{\phi}_\mu(f)$ is the same along different axes to achieve coherent addition of sensing field contributions. This is particularly important when performing AC sensing with interacting spin ensembles, as interaction decoupling requires the toggling-frame $\tilde{S}^z$ spin operator to spend equal time along each of the three axes (see Rule 2 above). For conventional interaction-decoupling pulse sequences that only utilize the effective sensing field along the $\hat{z}$-axis, the corresponding sensing efficiency can only be $1/3$ of that given by the XY-8 sequence. However, utilizing the full vector nature of the phase-synchronized effective sensing field $\vec{B}_\text{eff}$, as is achieved with Seq.~B, we can boost the sensitivity by a factor of $\sqrt{3}$, which is close to optimal for interaction-decoupling pulse sequences~\cite{otherpaper}.

\subsection{Detailed analysis of pulse sequences}
\label{sec:sequenceanalysis}
\begin{figure*}[h]
\begin{center}
\includegraphics[width=0.95\textwidth]{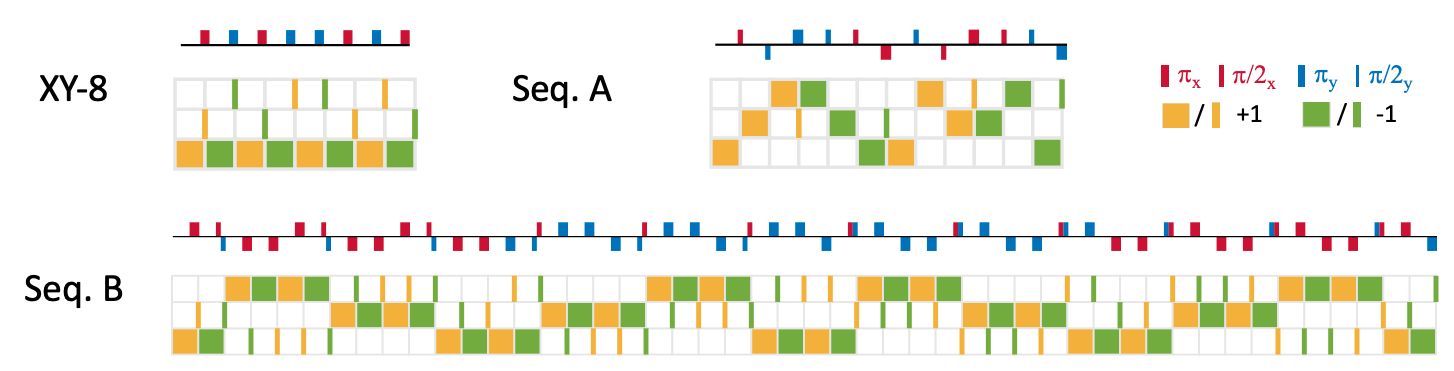}
\end{center}
\caption{Pictorial representation of pulse sequences introduced in the main text. The top row of each panel indicates the  conventional pulse sequence representation in terms of applied $\pi/2$ and $\pi$ pulses along the $\hat{x}$ and $\hat{y}$ axes, with above/below the black line indicating positive/negative rotations. The bottom row shows the frame matrix representation, illustrating the toggling-frame transformations of the $S^z$ spin operator; each row corresponds to a different axis direction, each column corresponds to a different free evolution period, and the filled boxes/bars indicate the spin operator orientations during free evolution periods/intermediate frames, with yellow/green indicating a positive/negative orientation. (\textbf{A}) XY-8 sequence, which decouples disorder but not interactions, (\textbf{B}) Sequence A, which decouples disorder and interactions in the infinitely short pulse limit, but does not address control imperfections arising from interactions acting during the finite pulse duration and rotation angle errors, (\textbf{C}) Sequence B, which suppresses disorder, interactions, and all finite pulse imperfections to leading order.}
\label{fig:seqrep}
\end{figure*}

We now apply the above rules for dynamical decoupling and fault-tolerance against leading-order imperfections to the sequences introduced in the main text. First, we introduce a convenient pictorial representation of the frame matrix $F_{\mu,k}$ (see Fig.~\ref{fig:seqrep}), which will help visualize the decoupling properties of the sequences~\cite{otherpaper}. Note that this representation is equivalent to that introduced in Fig.~3A of the main text, but has the advantage that the decoupling rules 1-4 can be readily analyzed for more complicated sequences. In each panel of Fig.~\ref{fig:seqrep}, the top row indicates the conventional representation of the corresponding sequence in terms of the pulses applied. The matrix below is a pictorial version of the frame matrix representation introduced above, where the three rows indicate the $\hat{x}$, $\hat{y}$, $\hat{z}$ axis directions and the columns indicate different free evolution periods of duration $\tau$. A yellow/green block in the $\mu$-th row, $k$-th column matrix indicates that $F_{\mu,k}=+1/-1$, i.e. the toggling frame spin operator $\tilde{S}^z_k=+S^\mu$/$-S^\mu$ during the $k$-th free evolution period. Meanwhile, the narrow bars in between the blocks indicate the spin operator orientation during the intermediate frames at the middle of $\pi$ pulses or composite $\pi/2$ pulses.

First, we analyze the XY-8 sequence, widely used for dynamical decoupling and AC sensing with non-interacting ensembles.
\begin{enumerate}
\item \green{\cmark} Disorder: There are an equal number of green and yellow squares/bars in each row of the matrix, indicating that disorder terms cancel for both free evolution periods and finite pulse durations.
\item \green{\cmark} Rotation angle errors: For each pair of axis directions, the total chirality of all rotations sum to zero. As an example, if we examine all interfaces between $\hat{y}$ and $\hat{z}$ frame directions, we see that the chirality of the first $\pi$ pulse ($+\hat{z}\rightarrow +\hat{y}\rightarrow -\hat{z}$) cancels with the third $\pi$ pulse ($+\hat{z}\rightarrow -\hat{y}\rightarrow -\hat{z}$), and similarly for the fifth and seventh pulses.
\item \red{\xmark} Interactions: All free evolution periods are located along the $\hat{z}$-axis (all filled squares are in the third row), so interactions are not fully symmetrized.
\end{enumerate}
Therefore, although the XY-8 sequence addresses on-site disorder and rotation angle errors, it does not suppress the effects of interactions, and consequently the observed coherence time is limited by spin-spin interactions.
\newline

Second, we analyze Seq.~A, which is designed to cancel the effects of disorder and interactions during the free evolution periods, but does not fully suppress all pulse-related imperfections.
\begin{enumerate}
\item \green{\cmark} Disorder: There are again an equal number of green and yellow squares/bars in each row of the matrix, indicating that disorder terms cancel for both free evolution periods and finite pulse durations.
\item \red{\xmark} Rotation angle errors: The chirality condition is not satisfied between the $\hat{y}$ and $\hat{z}$ directions; to see this, notice that the first $\pi/2$ pulse has chirality $-\hat{x}$, while the fifth and sixth pulses combine to also give a chirality $-\hat{x}$, such that the net chirality is not zero. This indicates that rotation angle errors are not fully suppressed.
\item \green{\cmark} Interactions during free evolution periods: There are an equal number of filled blocks along each row of the matrix, indicating that interaction terms are cancelled during free evolution periods.
\item \red{\xmark} Interactions from intermediate frames: There are more narrow bars in the $\hat{x}$ and $\hat{y}$ directions compared to the $\hat{z}$ direction, so the average interaction contribution in rule 2 is not cancelled for finite pulse durations.
\item \red{\xmark} Interaction cross-terms due to finite pulse durations: The parity condition is not satisfied between the $\hat{y}$ and $\hat{z}$ direction, since between these two rows there are three parity-preserving interfaces and one parity-changing interface. This indicates that the interaction cross-terms are also not fully suppressed for finite pulse durations.
\end{enumerate}
Therefore, while Seq.~A is expected to work well in the absence of pulse imperfections, the performance will be severely affected for finite pulse durations.
\newline

Finally, we apply the design rules to Seq.~B, and confirm that it satisfies all conditions for dynamical decoupling and fault-tolerance against leading-order imperfections.
\begin{enumerate}
\item \green{\cmark} Disorder: For the free evolution periods, the sequence shows a clear echo structure with alternating yellow and green squares, indicating that it should efficiently suppress disorder. The intermediate frames are also mostly directly paired into yellow and green pairs, except for the middle yellow bar along the $\hat{x}$ axis, which is paired with the last green bar along the same axis.
\item \green{\cmark} Rotation angle errors: The interface chirality sum for each pair of axes is zero, so rotation angle errors are fully cancelled. One simple way to see this without explicitly calculating the individual chiralities is to note that the $k$-th block/bar has opposite color from the $(n-k)$-th block/bar, where $n$ is the total number of blocks/bars, neglecting the final green bar in the $\hat{x}$-direction. Therefore, the chirality contributions at each interface will be precisely cancelled by another contribution from the opposite side of the sequence, summing to zero at the end.
\item \green{\cmark} Interactions during free evolution periods: There are an equal number (16) of filled square blocks along each axis direction, so interactions during free evolution periods are fully cancelled.
\item \green{\cmark} Interactions from intermediate frames:  There are an equal number (16) of narrow bars along each axis direction, so interactions during finite pulse durations are fully cancelled.
\item \green{\cmark} Interaction cross-terms due to finite pulse durations: The interface parity sum for each pair of axes is zero, so interaction cross-terms are fully cancelled. As an example, consider the interfaces between $\hat{y}$ and $\hat{z}$ axis directions: $\pi$ pulses have their parity contributions automatically cancelled, so we only need to consider the interfaces involved in the composite $\pi/2$ pulses that include an interface between $\hat{y}$ and $\hat{z}$. These are located after the following blocks (parity contribution indicated in parentheses): 2(+), 6(-), 18(+), 22(-), 26(-), 30(+), 42(-), 46(+), with pairwise cancellation of the signs.
\end{enumerate}

In addition to satisfying all of the above rules, Seq.~B is also designed with heuristics to suppress higher-order contributions. For example, it has a fast spin-echo structure for both free evolution periods and intermediate frames, such that the dominant disorder effects are echoed out as soon as possible. This is particularly important for our disorder-dominated spin ensemble, and will significantly reduce higher-order contributions to the effective Hamiltonian, since most of the commutators in higher-order terms that involve disorder will cancel out.

We now analyze the sensing properties of Seq.~B. As discussed in the preceding section, the sensitivity of a given pulse sequence to an AC external field at frequency $f$ and phase $\phi$ is characterized by the frequency-domain modulation function $|\tilde{F}_t(f)|$. This is plotted for Seq.~B in Fig.~3B of the main text, where the dominant resonance is highlighted. As can be seen in both Fig.~3A of the main text and Fig.~\ref{fig:seqrep}C, the AC signal will be well-synchronized with the frame changes along each axis in this case, and the signal phase $\phi$ for which the field strength along each axis is maximized coincides for different axes. This implies that the effective field direction will point along the $[1,1,1]$-direction, efficiently utilizing the phase accumulation along all three axes.

We note that using our algebraic rules, one can prove the necessity of composite $\pi/2$ pulse structures for robust interaction-decoupling sensing sequences that exhibit a fast spin-echo structure. This is because to effectively accumulate phase for this frequency $f$, the frame modulation pattern along each axis must be fixed (e.g. always appearing as $[+1,-1]$ along $\hat{x}$); this in turn implies that any interface between two axis directions will always have a fixed parity, leading to rule 3 above being violated. Thus, composite $\pi/2$ pulses, which can adjust the parity of frame-changing interfaces, must always be employed in such sequence design, as is the case with Seq.~B presented in the main text.

\subsection{Remaining limits to coherence}
Our sequence decouples disorder and interactions and suppresses dominant pulse-related imperfections at the leading-order average Hamiltonian level. These provide a significant extension of spin coherence times, with a 5-fold improvement over conventional dynamical decoupling sequences. However, the coherence times achieved here are not yet limited by the depolarization time $T_1$, which in the present sample is around 100 $\mu$s.

There are several factors that could limit the observed coherence times for Seq.~B: First, higher-order terms in the Floquet-Magnus expansion are not fully suppressed, leading to residual disorder and interaction effects. This could be particularly important for contributions involving the large on-site disorder in our electronic spin ensemble. Indeed, performing exact diagonalization simulations for different disorder strengths, we find that the coherence time significantly extends as one reduces the disorder strength. Second, microwave waveform imperfections due to e.g. interface reflections, could affect the efficiency of dynamical decoupling and reduce coherence times. Indeed, we have found that when the attenuators along the microwave path are removed, the larger reflections significantly reduce the observed coherence times and cause oscillations in the signal. While the attenuators alleviate this issue, there could still be residual effects. Third, there may be small residual contributions such as {\it time-dependent} disorder effects from spin-bath dynamics of the environment. Finally, imperfect spin polarization into the $|m_s=0\rangle$ state could cause additional spin-exchange dynamics out of the two-level system formed by $|m_s=0,-1\rangle$, causing decoherence. However, we find that intentionally moving part of the spin population into the $|m_s=+1\rangle$ state during state preparation does not significantly affect the coherence time, indicating that this decay channel is probably not dominant. Note that imperfect polarization within the $|m_s=0,-1\rangle$ spin levels is not expected to have an impact on the coherence times; only the polarization into $|m_s=+1\rangle$ should have an effect.

\section{AC magnetic-field sensing with interacting spins}
\subsection{Calibration of AC magnetic field amplitude}
\label{sec:fieldcalibration}
\begin{figure*}[h!]
\begin{center}
\includegraphics[width=0.95\textwidth]{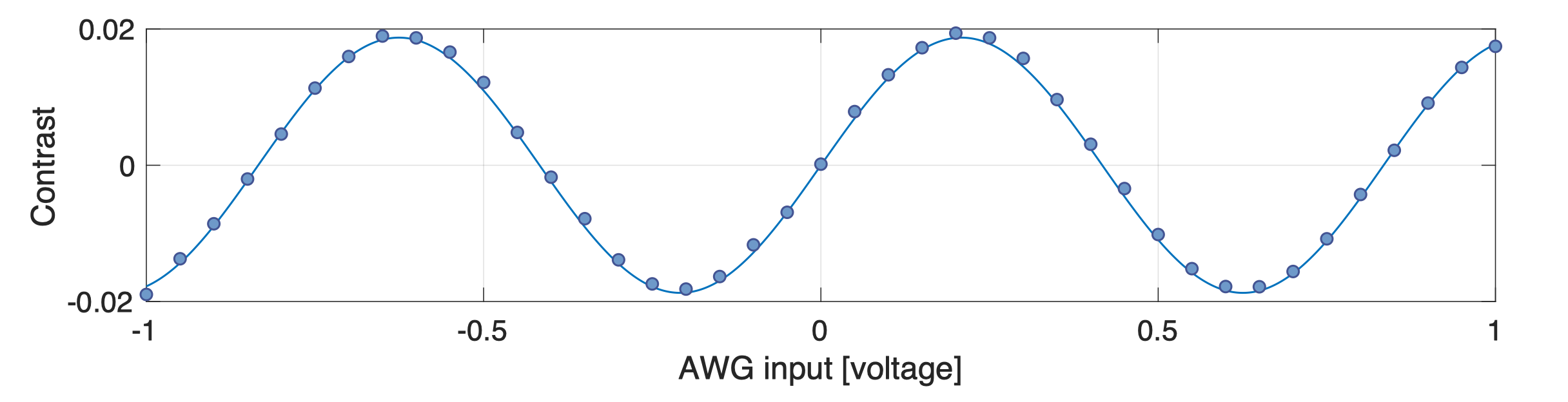}
\end{center}

\caption{Calibration of sensing signal amplitude. A conventional XY-8 sequence is employed to perform AC magnetometry and calibrate the magnitude of the AC magnetic field. A continuous-wave signal is directly synthesized by the AWG and optimally synchronized with the XY-8 sequence to maximize phase accumulation. To satisfy the resonance condition, the frequency of the AC signal is set to $f = \frac{1}{2(\tau+t_\pi)}$, where $\tau$ is the free evolution interval between adjacent $\pi$ pulses of the XY-8 sequence and $t_\pi$ is the $\pi$ pulse length. Here, we fix the phase accumulation time to $t = 2.16~\mu$s. The spin contrast is monitored as a function of AWG input voltage. The measured trace is fitted to a sinusoidal curve (solid line) to extract the frequency of the contrast oscillation. Using the extracted frequency, we obtain a calibration constant of 55.9(1)~$\mu$T/V.}
\label{fig:XY8calibration}
\end{figure*}

In the following subsections, we provide details on the responses of our NV-ensemble magnetometer. To characterize the performance of AC magnetic field sensing, a continuous-wave sinusoidal signal is generated by the AWG with a controlled amplitude and phase relative to the NV sensing sequence. To calibrate the strength of the magnetic field, we employ an XY-8 sequence and measure the contrast $S$ as a function of AWG input voltage (Fig.~\ref{fig:XY8calibration}). The contrast $S$ shows periodic oscillations with increasing input amplitude, indicating that the NV centers undergo a coherent precession due to the external AC magnetic field. Analytically, the contrast oscillation can be fitted to 
\begin{equation}
S = S_{\text{max}} \sin\left(\frac{2\gamma_{NV} \beta B_\text{AC} t}{\pi}\right),
\end{equation}
where $S_{\text{max}}$ is the maximal signal contrast, $\gamma_{\text{NV}} = (2\pi)$28~GHz/T is the gyromagnetic ratio of the NV center, $t$ is the phase accumulation time for sensing, $\beta = 0.95$ is a small correction factor due to the finite duration of $\pi$ pulses used in the XY-8 sequence, and $B_\text{AC}$ is the AC magnetic field amplitude~\cite{degen2017quantum}. A comparison of the extracted $B_\text{AC}$ with the applied voltage allows the precise calibration of the AC magnetic field generated for a given AWG output voltage. The results are shown in Fig.~\ref{fig:XY8calibration}, where we obtain a calibration constant of 55.9(1)~$\mu$T/V.

\subsection{Sensing response as a function of relative phase between pulse sequence and signal}
\label{sec:relphase}
\begin{figure*}[h!]
\begin{center}
\includegraphics[width=0.9\textwidth]{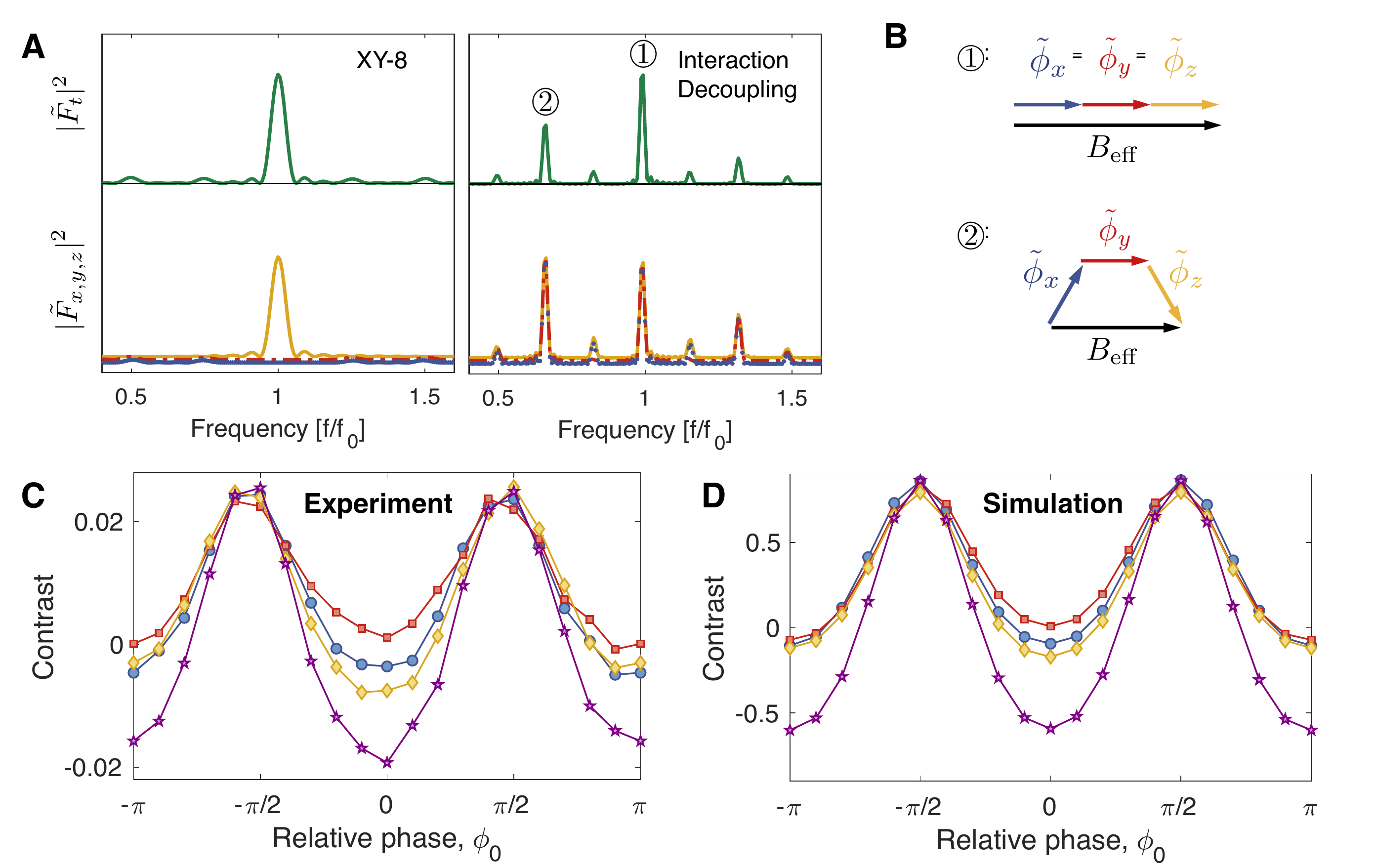}
\end{center}
\caption{Dependence of NV-ensemble magnetometer response on the relative phase delay between the sensing signal and sequence. \textbf{(A)} Resonance spectra for XY-8 (left) and an interaction-decoupling sequence (right), including their total spectral intensities (green) and individual spectral intensities for each axis $\hat{x}$ (blue), $\hat{y}$ (red) and $\hat{z}$ (yellow). \textbf{(B)} Illustration of three-axis spectral-phase synchronization at resonance \textcircctr{1}, and out-of-synchrony at resonance \textcircctr{2}, resulting in stronger and weaker effective magnetic field strengths (black arrows), respectively. \textbf{(C)} Experimental data and \textbf{(D)} numerical simulations for Seq.~B. Different colors denote different initialization conditions, corresponding to initial spin orientations along the $\hat{x}$ (blue circles), $\hat{y}$ (red squares), $\hat{z}$ (yellow diamonds), and the optimal $[1,1,-2]$ (purple stars) axes. In (\textbf{D}), we numerically solve the time-dependent Schr$\ddot{\text{o}}$dinger equation for the full many-body Hamiltonian for 6 spins. We use a time step $dt=1$ ns for numerical integration. The sensor spins are randomly generated in 3D using estimated experimental NV concentrations, with periodic boundary conditions. Each spin has a random Gaussian on-site disorder with standard deviation $\sim(2\pi)~4.0$~MHz. Long-range dipolar interactions ($\sim1/r^3$) are assumed with a coupling strength determined by the relative distance and orientation between the spins. The simulation parameters are chosen in accordance with the experiments ($\tau$ = 25 ns, $t_{\pi}$ = 20 ns, $t_{\pi/2}$ = 10 ns, $t$ = 6.516~$\mu$s, $N=3$ Floquet periods, $T_1$ = 100~$\mu$s with stretched exponential of power 0.5, and the frequency/amplitude of the AC signal are set to 11 MHz and 6.6~$\mu$T).}
\label{fig:phasesweep}
\end{figure*}

We now consider the sensing response as a function of the relative phase between the pulse sequence and signal. First, we consider the relative spectral phase alignment between different axes, $\tilde{\phi}_x(f),\tilde{\phi}_y(f),\tilde{\phi}_z(f)$. In conventional sensing sequences such as XY-8, only the effective field along a single axis is employed for sensing (Fig.~\ref{fig:phasesweep}A, left), and hence the relative phase alignment between axes is not important. In the case of interaction-decoupling sequences, however, the effective field along all three axes contribute to the signal, and hence spectral phase alignment between the different axes is crucial to achieve highest sensitivity. To illustrate this, consider the spectral response of an interaction decoupling sequence that sequentially transforms the $S^z$ operator into $+S^z,-S^z,+S^y,-S^y,+S^x,-S^x,+S^z$ etc., as shown in Fig.~\ref{fig:phasesweep}A. It contains two pronounced peaks located at $f = (2/3)f_0$ and $f = f_0$, which we label as resonance \textcircctr{2} and \textcircctr{1}, respectively (Fig.~\ref{fig:phasesweep}A, right). Examining the individual strengths of the spectral functions $|\tilde{F}_{x,y,z}|$, resonance \textcircctr{2} has a higher peak intensity than resonance \textcircctr{1}. However, if one examines the total spectral strength $|\tilde{F}_t|$, resonance \textcircctr{1} shows a significantly higher peak, giving maximum sensitivity (Fig.~\ref{fig:phasesweep}A, top right). This is because at resonance \textcircctr{1}, all three spectral phases are aligned ($\tilde{\phi}_x=\tilde{\phi}_y=\tilde{\phi}_z$, Fig.~\ref{fig:phasesweep}B, top) and thus generate the largest effective sensing field along the [1,1,1]-direction, while for resonance \textcircctr{2}, the asynchronous phases lead to destructive field addition and hence a smaller total field strength (Fig.~\ref{fig:phasesweep}B, bottom).

In the following, we thus assume that the spectral phases at resonance frequency $f$ are aligned between different axes, giving $\tilde{\phi}_x(f) = \tilde{\phi}_y(f) = \tilde{\phi}_z(f)$, as is the case for our sensing sequence Seq.~B. In this case, the phase accumulation of the sensor due to an external signal is constructive if the signal and applied pulse sequence are in phase, i.e., when the relative phase $\phi_0$, defined as $\phi_0 = \phi - \tilde{\phi}_{z}(f)$, between the two is 0 or $\pm\pi$. If $\phi_0 = \pm\pi/2$, then there will be zero phase accumulation. To experimentally confirm this relative phase dependence, we sweep the value of $\phi_0$ from $-\pi$ to $+\pi$ using Seq.~B and measure contrast variations as a function of signal phase (see Fig.~\ref{fig:phasesweep}C). Interestingly, the phase responses for the $\hat{x}$, $\hat{y}$ and $\hat{z}$-axis initial states are not identical to each other, although the effective field strength along each axis is expected to be the same at the average Hamiltonian level. The observed asymmetries are attributed to the fact that at finite target sensing signal amplitudes, the rotations induced by the sensing signal at different times do not commute with each other (disorder-dominated higher-order contributions), and thus the dynamics exhibit deviations from the average Hamiltonian picture and the symmetry between different axis directions is broken.

Despite this non-trivial phase dependence, an optimal state preparation orthogonal to the $[1,1,1]$-direction still shows largest contrast and results in maximal sensitivity. In addition, we note that the sign-inversion symmetry of the signal field is broken: The sensor response at $\phi_0 = 0 $ is not the same as that of $\phi_0 = \pi$, likely due to the aforementioned non-commuting effect. To lend support to this, a numerical simulation is performed by solving the time-dependent Schr$\ddot{\text{o}}$dinger equation in the presence of a target AC sensing field, where we include the on-site disorder and interactions in the system. The result is shown in Fig.~\ref{fig:phasesweep}D, showing qualitatively similar behavior to the experimental data and suggesting that this effect indeed originates from the non-commuting nature of $\pi/2$ rotations in the presence of an external sensing field.

We now rigorously derive the optimal signal phase at which the contrast is maximized, for a given effective magnetic field and initial state. This will allow us to model the electron spin resonance (ESR) frequency response in the general case of {\it vectorial} effective magnetic fields in the toggling frame, going beyond the conventional modeling of ESR responses that only consider $\hat{z}$-directional sensing fields.

To be more concrete, we consider the evolution of a given initial state $|\psi_0\rangle$ for phase accumulation time $t$ under some pulse sequence and target AC magnetic field, with the Hamiltonian described in Sec.~\ref{sec:sequencedesign}. In the toggling frame, the dynamics of the spin corresponds to a precession of angle $\theta=\gamma tB_t/\hbar$ around the axis $(u_x,u_y,u_z) = (B_x,B_y,B_z)/B_t$, with the total magnetic field $B_t = \sqrt{B_x^2 + B_y^2 + B_z^2}$. Here, the individual field components already take into account the relative phase alignment with respect to the target signal field, namely, $B_\mu= B_\text{AC} |\tilde{F}_\mu(f)| \Re[e^{-i(\phi_\mu(f)-\phi)}]$ for $\mu=x,y,z.$ Writing the initial state as a vector on the Bloch sphere, the evolution corresponds to the rotation matrix
\begin{align}
R = \left[ \begin{array}{ccc}{\cos \theta+u_{x}^{2}(1-\cos \theta)} & {u_{x} u_{y}(1-\cos \theta)-u_{z} \sin \theta} & {u_{x} u_{z}(1-\cos \theta)+u_{y} \sin \theta} \\ {u_{y} u_{x}(1-\cos \theta)+u_{z} \sin \theta} & {\cos \theta+u_{y}^{2}(1-\cos \theta)} & {u_{y} u_{z}(1-\cos \theta)-u_{x} \sin\theta} \\ {u_{z} u_{x}(1-\cos \theta)-u_{y} \sin \theta} & {u_{z} u_{y}(1-\cos \theta)+u_{x} \sin \theta} & {\cos \theta+u_{z}^{2}(1-\cos \theta)}\end{array}\right].
\label{eq:rotmat}
\end{align}
For cosine magnetometry, the observed contrast then corresponds to the deviation from unity of the projection of the rotated state onto the initial state. As discussed above, in the experiments, we optimize the relative phase delay between the sensing signal and the pulse sequence to maximize the contrast. For numerical comparisons, it is therefore important to determine the optimal signal phase for which the contrast is maximized.

In the following, we show that for small signal field rotations, the contrast is maximized when the phase is chosen such that the field strength orthogonal to the initialization direction is maximized, regardless of the field strength along the initialization direction. To illustrate this, we first consider the case of spins initialized along the $\hat{x}$-direction. After a phase accumulation time $t$, the projection onto the initialization direction is given by the (1,1)-element of the rotation matrix in Eq.~(\ref{eq:rotmat}). Expanding for small $\theta$, we find the contrast to be
\begin{align}
C=1-\qty[\cos \theta+u_{x}^{2}(1-\cos \theta)]\approx \frac{\theta^2}{2}\qty(1-u_x^2)=\frac{\gamma^2t^2}{2\hbar^2}\qty(B_y^2+B_z^2).
\end{align}
Thus, to maximize contrast, we need to maximize the total field strength in the $\hat{y}$ and $\hat{z}$ directions. For a generic initialization direction, we can redefine the coordinate system to have one of the basis vectors pointing along the chosen initialization direction, and rewrite the effective magnetic field $\vec{B}_\text{eff} = [B_x, B_y, B_z]$ in this basis. For example, for the optimal initialization direction $[1,1,-2]$ for Seq.~B in the main text, we can choose basis vectors $[1,1,-2]/\sqrt{6}$, $[1,1,1]/\sqrt{3}$, $[1,-1,0]/\sqrt{2}$. The signal contrast is then maximized when the projection of $\vec{B}_\text{eff}$ onto the latter two transverse directions is maximal, which can be easily written down based on linearity of the field component addition.

Combining these results with the calculated frequency-domain modulation functions $\tilde{F}_\mu(f)$, we can easily choose the phase of the AC magnetic field signal that gives maximal contrast under each measurement condition, and use this information to determine the expected contrast. This allows us to generate the theoretical ESR curves in Fig.~3D of the main text.

\subsection{Initialization and readout protocols}
Similar to the conventional sine and cosine magnetometry protocols, for our new sequences, we can also design spin initialization and readout protocols to maximize the sensitivity to overall polarization (sine magnetometry) or fluctuations in polarization (cosine magnetometry) at close to zero signal field. Both of these protocols require that the spin is initialized in a direction perpendicular to the effective magnetic field direction in order to maximize precession.

In our experiment, we vary the initialization direction to implement sine and cosine magnetometry, while fixing the readout rotation. For readout, we rotate around an axis perpendicular to the effective magnetic field direction, and choose the rotation angle to rotate the precession plane to contain the $\hat{z}$-axis for state measurement via spin-state-dependent fluorescence. Here, we choose to rotate the $\vec{v}_1=[1,1,1]/\sqrt{3}$ effective field direction into $\vec{v}_2=[1,1,0]/\sqrt{2}$, so that the precession plane perpendicular to the effective field direction contains the $\hat{z}$-axis. This corresponds to a rotation around the $[-1,1,0]$-axis by an angle $\arccos(\vec{v}_1\cdot\vec{v}_2)=\arccos(\sqrt{2/3})$.

With the readout rotation determined, one can now easily find the initialization rotations that are analogous to cosine and sine magnetometry. More specifically, cosine magnetometry corresponds to initializing the spins at the location of maximum contrast, while sine magnetometry corresponds to initializating the spins at the location of zero contrast. In our experiments, we realize cosine magnetomery (Fig.~3D of the main text) by a rotation of angle $\arccos(\sqrt{2/3})$  around the $[1,-1,0]$ axis, to rotate into the state $[-1,-1,2]/\sqrt{6}$, and realize sine magnetometry (Fig.~4 of the main text) by a rotation of angle $\pi/2$ around the $[1,1,0]$ axis, to rotate into the state $[1,-1,0]/\sqrt{2}$. Under the readout pulse, the former is rotated to the $[0,0,1]$ state (maximal contrast), while the latter is not rotated and remains on the equator (zero contrast). Note that this is slightly different from the conventional approach for switching between sine and cosine magnetometry, where one fixes the initialization rotation and changes the readout rotation, but the two methods are equivalent to each other. For example, we can choose instead to always rotate the spin into the state $[1,-1,0]/\sqrt{2}$ for initialization. The readout rotation for sine magnetometry remains unchanged, while we can choose a $\pi/2$ rotation around the $[1,1,0]$ direction for cosine magnetometry. Note however, that this requires a rotation of a larger angle than our protocol; in fact, our initialization and readout protocol is the one that requires minimal rotation angle.

Based on the above discussions, we can also easily evaluate the effect of choosing optimal initialization and readout directions compared to the conventional initialization and readout direction $\hat{x}$. While the optimal initialization direction achieves full contrast, an $\hat{x}$-axis initialization is not perpendicular to the precession axis (the tilt angle is instead $\alpha=\arccos(1/\sqrt{3})$), producing a contrast that is a fraction $[1-\cos(2\alpha)]/2=2/3$ of the full contrast.

\subsection{Sensing response as a function of sensing signal amplitude}
\begin{figure*}[h]
\begin{center}
\includegraphics[width=0.9\textwidth]{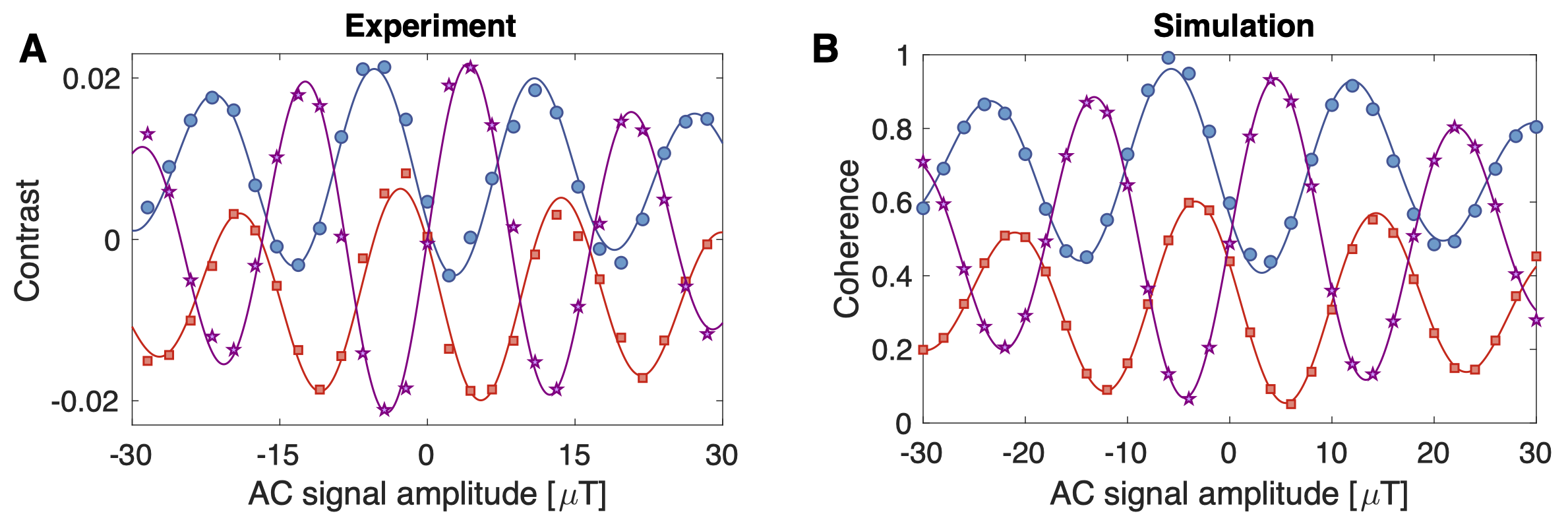}
\end{center}
\caption{Dependence of NV-ensemble magnetometer response on AC signal amplitude under the leading-order fault-tolerant sensing sequence (Seq.~B). \textbf{(A)} Experimental data and \textbf{(B)} numerical simulation. In (\textbf{A}), different markers denote different measurement conditions, each corresponding to initial spin orientations along the $\hat{x}$ (blue circles), $\hat{y}$ (red squares), and optimal $[1,-1,0]$ (purple stars) axes, while different lines indicate a fit taking the form of a Gaussian-decaying sinusoidal function with an offset. In (\textbf{B}), we numerically solve the time-dependent Schr$\ddot{\text{o}}$dinger equation for the full many-body Hamiltonian consisting of 6 spins to capture interaction effects. The same simulation parameters are used as in Fig.~S7.}
\label{fig:ampsweep}
\end{figure*}

In order to evaluate the sensitivity of our NV-ensemble magnetometer, the amplitude of a continuous-wave signal is swept while the relative phase between the pulse sequence and signal is locked to the optimal phase delay $\phi_0 = 0$. With our new sensing sequence Seq.~B, we monitor how the contrast $S$ varies as a function of signal amplitude at a fixed phase accumulation time. As shown in Fig.~\ref{fig:ampsweep}A, the contrast exhibits sinusoidal oscillations for different initial states in response to the AC sensing signal. However, due to the effective precession axis along the $[1,1,1]$ direction under the pulse sequence, spin initializations along the $\hat{x}$- and $\hat{y}$-axes result in asymmetric oscillations with reduced contrast, introducing positive and negative offsets to $S$, respectively. On the other hand, if spins are prepared along the $[1,-1,0]$ direction and rotated by angle $\arccos(\sqrt{2/3})$ around $[-1,1,0]$ for readout (sine magnetometry explained above), then the contrast oscillation remains symmetric with zero offset, with an increased slope at the zero-field point (see Fig.~\ref{fig:ampsweep}A). This unconventional state initialization and readout is crucial to achieve optimal sensitivity.

As the AC signal amplitude is further increased, we find that the peak-to-peak contrast starts to decrease. As shown in Fig.~\ref{fig:ampsweep}B, a qualitatively similar behavior is also observed in a numerical simulation where the time-dependent many-body dynamics is solved for a dipolar spin system subject to the same dynamical decoupling pulse sequence and an external AC signal. We attribute the reduction in constrast to a decrease in disorder and interaction decoupling efficiency of the pulse sequence, due to the presence of a large AC signal that modifies the effective spin frame significantly, causing the average Hamiltonian picture in the frames specified by the pulses to partially break down~\cite{cory1996distortions}. Numerically, we find that this effect is dominated by contributions from the reduced disorder decoupling efficiency. To account for these effects, in Fig.~\ref{fig:ampsweep}, we fit the contrast oscillations to a sinusoidal curve with a Gaussian decay.

\section{Sensitivity estimation}
\label{sec:sensitivity}
\subsection{AC magnetic field sensitivity}
The AC magnetic-field sensitivity $\eta$ is defined as the minimum detectable signal that yields unit SNR for an integration time of 1 second \cite{taylor2008high}, and can be estimated by
\begin{equation}
\eta = \frac{b\pi e^{\qty(t/T_2)^\alpha}}{2 \gamma_{\text{NV}} C \sqrt{K}}\frac{\sqrt{t + t_m}}{t},
\label{eq:sensitivity}
\end{equation}
where $t$ is the phase accumulation time, $T_2$ is the coherence time of the ensemble, $\alpha$ is the exponent of the stretched exponential decoherence profile, $C$ is an overall readout efficiency parameter, $K$ is the number of NV centers in a confocal volume, $t_m$ is the additional time needed to initialize and read out the NV centers and $b$ is a pulse-sequence-dependent sensitivity factor~\cite{degen2017quantum}. For example, $b = 1$ for the XY-8 sequence and $b=\sqrt{3}$ for Seq.~B. Assuming a single group sensor density of 4 ppm and a sensor volume of $V = 8.1 \times 10^{-3}~\mu \text{m}^3$, we identify $K \approx 6 \times 10^3$ spins. By fitting Eq.~(\ref{eq:sensitivity}) to the measured sensitivity curves presented in Fig.~4B of the main text, we estimate the overall readout efficiency per NV as $C \approx 2.8 \times 10^{-3}$. Note that in the curves in Fig.~4B of the main text, the coherence decay parameters $T_2$ and $\alpha$ are chosen to be free fitting parameters constrained to be close to the independently-determined values in Fig.~2 of the main text. A small discrepancy of the fit parameters from the measured ones is attributed to the effects of charge dynamics under varying green laser duty cycles, which can slightly modify the observed sensitivity values.

\subsection{Sensitivity optimization}
\label{sec:sensopt}
The sensitivity can be experimentally quantified by
\begin{equation}
\eta = \frac{\sigma_S^\text{1s}}{|dS/dB_\text{AC}|}
\label{eq:eta}
\end{equation}
where $\sigma_S^\text{1s}$ is the uncertainty of the contrast $S$ for 1 second averaging and $|dS/dB_\text{AC}|$ is the gradient of $S$ with respect to the field amplitude $B_\text{AC}$~\cite{degen2017quantum}. The uncertainty $\sigma_S^\text{1s}$ and the slope $|dS/dB_\text{AC}|$ are extracted at the zero-field point ($B_\text{AC}=0$), from which we determine the maximum sensitivity (see Fig.~4A in the main text). The contrast $S$ is proportional to the probability $p$ since $S$ measures the population difference between the $\ket{0}$ and $\ket{-1}$ states.

\begin{figure*}[h]
\begin{center}
\includegraphics[width=0.8\textwidth]{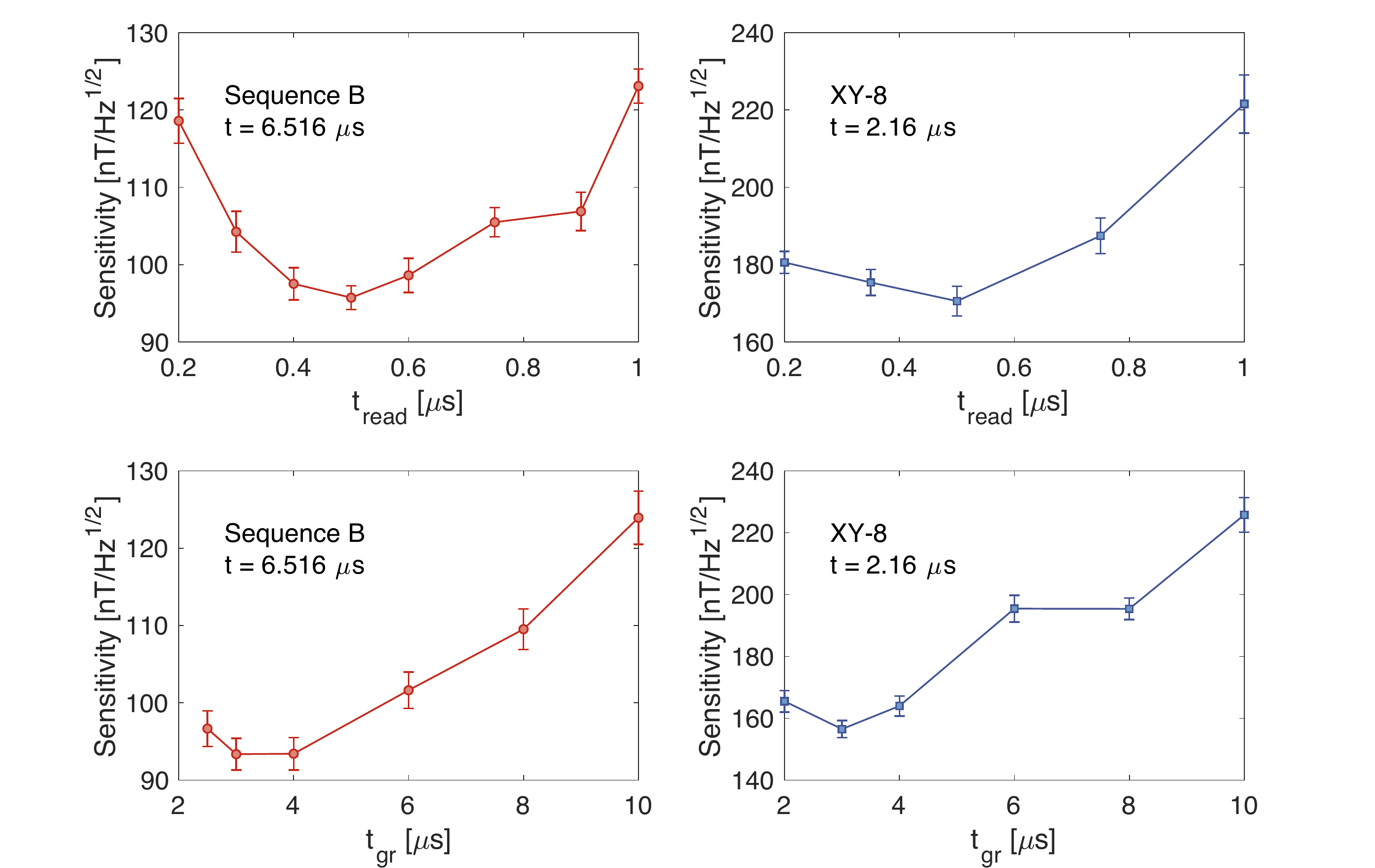}
\end{center}
\caption{Optimization of sensitivity. To identify the parameter regime where sensitivity is optimized, the counter length $t_\text{read}$ (\textbf{A,B}) and green duration $t_\text{gr}$ (\textbf{C,D}) are varied to find the optimal point balancing readout signal contrast and readout integration uncertainty. We find that $t_\text{read} =500~$ns and $t_\text{gr} = 4~\mu$s show the highest sensitivity for both Seq.~B (\textbf{A,C}) and XY-8 (\textbf{B,D}). The errorbars denote the standard deviations of sensitivity values, obtained via error propagation.}
\label{fig:sensopt}
\end{figure*}

In order to optimize the sensitivity, NV state preparation and readout parameters, including counter length $t_{\text{read}}$ and green duration $t_{\text{gr}}$, are fine-tuned. In our experiments, the measurement and readout overhead time is $t_m = t_{\text{gr}} + 0.3~\mu$s (including a short buffer to adjust for time delays between different equipment) and the sensor readout is performed at the leading edge inside the green illumination. As shown in Fig.~\ref{fig:sensopt}, $t_{\text{read}}=500~$ns and $t_\text{gr} =4~\mu$s result in the best sensitivity values for both Seq.~B and XY-8. With these parameters, we probe the sensitivity of our dense NV ensemble as a function of phase accumulation time. As presented in Fig.~4B in the main text, the sensitivity scaling calculated from Eq.~(\ref{eq:sensitivity}) is in excellent agreement with the experimental observations. The minimum sensitivity currently achieved in our system is estimated to be 92(2)~nT/$\sqrt{\text{Hz}}$ measured under Seq.~B, which is a factor of $\sim$1.7 improved over the conventional XY-8 sequence. This translates into a volume-normalized sensitivity of 8.3(8)~nT$\cdot\mu$m$^{3/2}/\sqrt{\text{Hz}}$, which is among the best volume-normalized sensitivities demonstrated using electronic spin ensembles, see Tab.~\ref{fig:senscompare}.

\subsection{Robustness against rotation-angle errors}
\begin{figure*}[h]
\begin{center}
\includegraphics[width=0.5\textwidth]{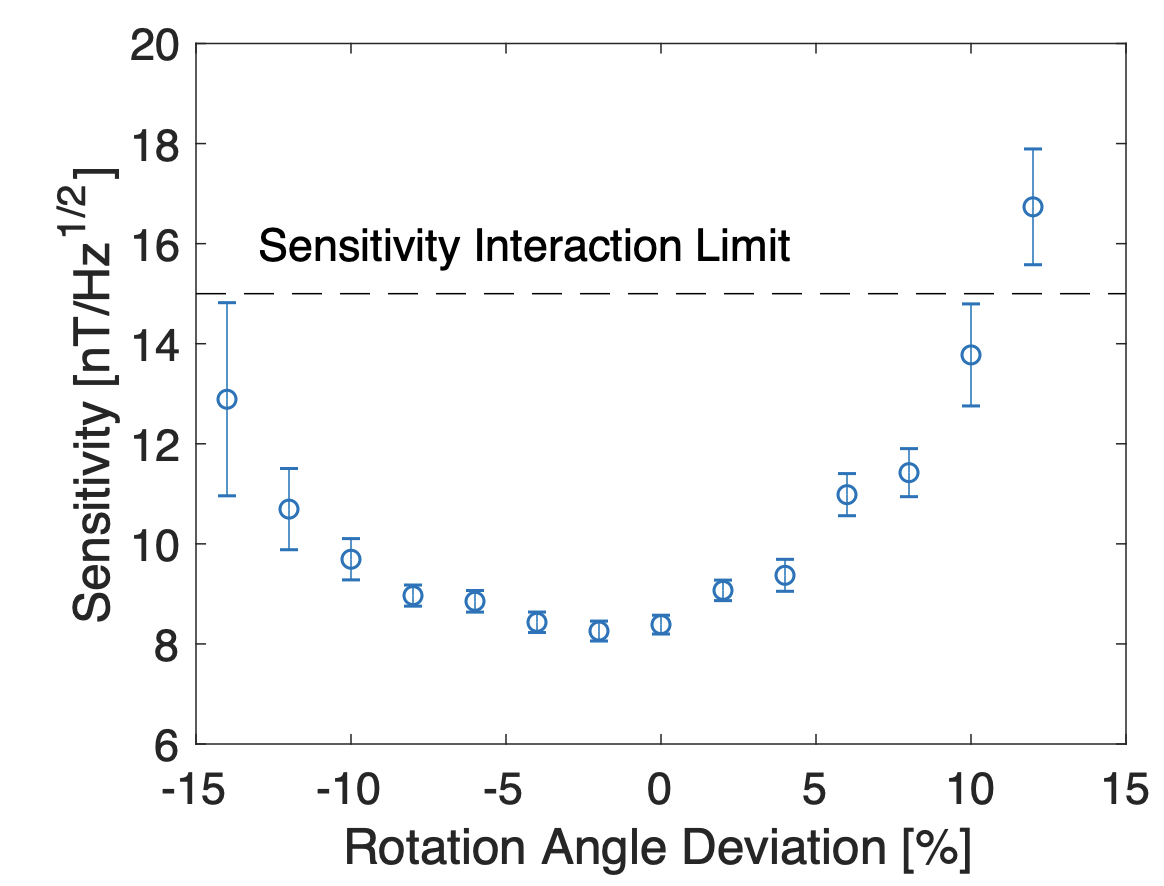}
\end{center}
\caption{Robustness of magnetic field sensitivity against rotation angle errors. The measured sensitivity is found to surpass the interaction limit for a wide range of systematic rotation angle deviations, up to $\sim12\%$.}
\label{fig:anglesweep}
\end{figure*}

We now explore the effect of systematic rotation angle errors on the sensitivity; this is particularly relevant for spin ensembles in a large probing volume, where the driving microwave field can have a sizable field-strength variation in space. As shown in Fig.~\ref{fig:anglesweep}, we find that the resulting sensitivity under Seq.~B is insensitive to a rotation angle deviation of up to $\sim\pm12\%$, exhibiting very good robustness against rotation angle errors. For this wide range of errors, Seq.~B still surpasses the interaction limit and shows a sensitivity that is better than a perfect XY-8 sensing sequence. This suggests that these sequences will also be very effective in applications where there is a sizable driving field inhomogeneity.

\subsection{Comparison of spin-ensemble-based sensors}

The figure of merit for ensemble-based sensors is the volume-normalized sensitivity. As shown in Tab.~\ref{fig:senscompare}, we compare our volume-normalized sensitivity with existing work on 3D ensemble magnetic field sensing. We calculate the overall readout efficiency parameter $C$ for each system using Eq.~(\ref{eq:sensitivity}) and values extracted from Refs.~\onlinecite{Pham2012Enhanced,Le2012Efficient,Wolf2015Subpicotesla,farfurnik2018spin,masuyama2018extending}. Despite the relatively low $C$ value extracted for our system, our dense spin ensemble achieves volume-normalized sensitivities among the best reported so far~\cite{footnote}. 

While our work is the first to push the sensitivity below the limit imposed by interactions, this advance may not be obvious from a direct comparison of the $\eta_V$ values. In addition to the differences in $C$, which arise mainly from different photon collection strategies, we attribute this fact to a sizable uncertainty in the quoted spin densities for NV-based diamond sensors. In principle, for XY sequences, the $\rho\cdot T_2$ product (the figure of merit governing the sensitivity scaling) is expected to lie below a universal interaction limit, given by the upper limit of $T_2$ imposed by interactions, where $\rho \cdot T_2$ reaches a constant number for any spin density~\cite{mitchell2019sensor}. The quoted $\rho\cdot T_2$ values, however, display large variations, with some a few orders of magnitude larger than our value, indicating that the high spin densities claimed in some works may have been over-estimated~\cite{masuyama2018extending}. This also leads to an under-estimate of the readout efficiency parameter $C$. In particular, at high densities, due to charge dynamics, the {\it effective} NV center density (as extracted in Sec.~S1 above) is expected to be lower than the value simply estimated from the conversion efficiency of nitrogen defects to NV centers. Moreover, extracting the effective NV center density from numerical simulations can be a non-trivial task at these high densities, since it is challenging to accurately describe long-range interacting spin dynamics using only a limited system size.

\renewcommand{\figurename}{Table}
\setcounter{figure}{0}
\begin{figure*}[h]
\begin{center}
\includegraphics[width=1\textwidth]{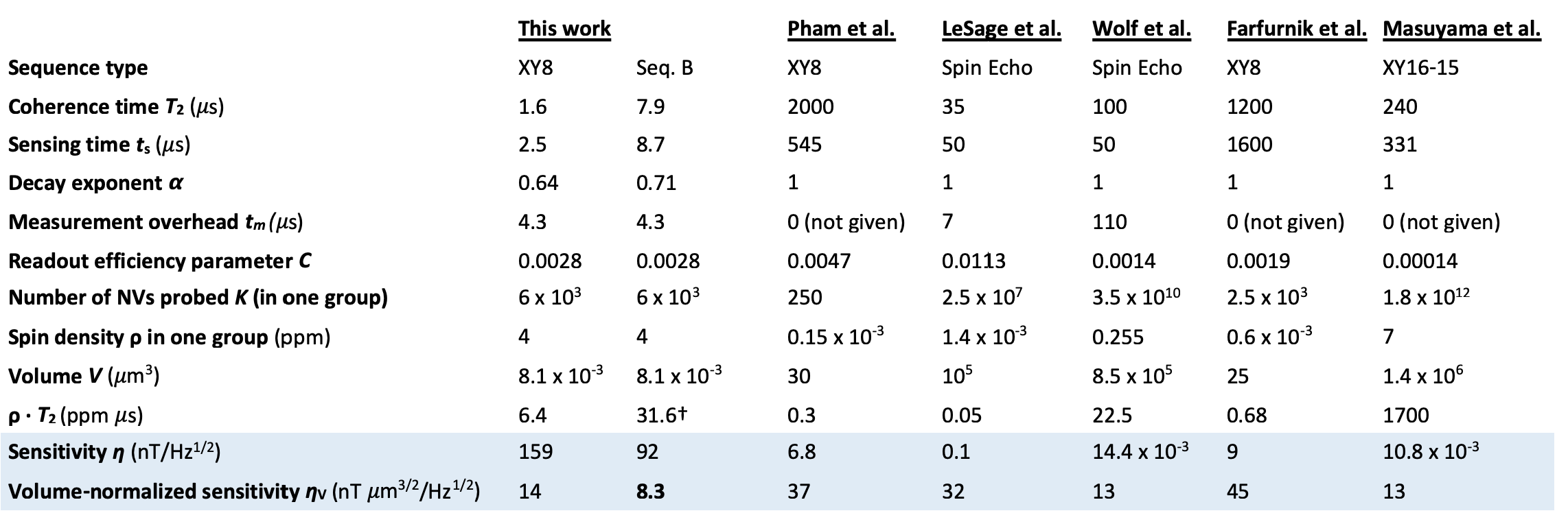}
\end{center}
\caption{Comparison of magnetometer performance for ensemble spin sensors with different spin densities. $^\dagger$: our experiments with Seq.~B operate beyond the interaction limit.}
\label{fig:senscompare}
\end{figure*}
\renewcommand{\figurename}{FIG.}
\setcounter{figure}{10}

We expect that the current sensitivity can be further enhanced by optimizing sensor characteristics and readout efficiencies. To this end, we estimate the volume-normalized sensitivity to improve up to single-digit picotesla levels of $\eta_\mathrm{V} = 8$~pT$\cdot\mu$m$^{3/2}/\sqrt{\text{Hz}}$ attainable under our phase-synchronized sensing sequence (Seq.~B). Here, we used the realistic system parameters of lifetime-limited coherence times of $T_2 = 2 T_1 = 200\,\mu$s, a total spin density of 15 ppm and $C = 0.2$~\cite{edmonds2012production,fukui2014perfect,michl2014perfect,lesik2015preferential,miyazaki2014atomistic}. For the projected $C$ we assume a collection efficiency above 5\%, which can be achieved through nanofabrication techniques such as nanoscale parabolic reflectors~\cite{taylor2008high,wan2018efficient}.

\bibliography{SensingExp_supp_bibtex}